\documentclass[twocolumn,showpacs,showkeys,amssymb,nobibnotes,superscriptaddress,aps,pra,bibnotes]{revtex4-1}

\usepackage{graphicx}
\usepackage{dcolumn}
\usepackage{bm}
\usepackage[T1]{fontenc}
\usepackage{mathptmx}
\usepackage{mathrsfs}
\usepackage{physics}
\usepackage{simplewick}
\usepackage{mathrsfs}
\usepackage{amsthm,amsmath,amssymb}

\usepackage[bottom]{footmisc}

\usepackage{amssymb}
\usepackage{amsmath}
\usepackage{amsfonts}
\usepackage{bm}
\usepackage{graphicx}
\usepackage{subfigure}
\usepackage[dvipsnames]{xcolor}
\usepackage{calc}
\usepackage{epsfig}
\usepackage{color}
\usepackage[colorlinks,citecolor=blue]{hyperref}
\begin{document}
\bibliographystyle{apsrev4-1}
\newcommand{\be}{\begin{equation}}
\newcommand{\ee}{\end{equation}}
\newcommand{\bs}{\begin{split}}
\newcommand{\es}{\end{split}}
\newcommand{\R}[1]{\textcolor{red}{#1}}
\newcommand{\B}[1]{\textcolor{blue}{#1}}

\title{Quantum scrambling in a toy model of photodetectors}
\author{Yubao Liu}
\affiliation{Center for Gravitational Experiment, Hubei Key Laboratory of Gravitation and Quantum Physics, School of Physics, Huazhong University of Science and Technology, Wuhan, 430074, China}
\author{Haixing Miao}
\email{haixing@tsinghua.edu.cn}
\affiliation{State Key Laboratory of Low Dimensional Quantum Physics, Department of Physics, Tsinghua University, Beijing, China}
\author{Yanbei Chen}
\email{yanbei@caltech.edu}
\affiliation{Burke Institute of Theoretical Physics, California Institute of Technology, Pasadena, CA, 91125, USA}
\author{Yiqiu Ma}
\email{myqphy@hust.edu.cn}
\affiliation{Center for Gravitational Experiment, Hubei Key Laboratory of Gravitation and Quantum Physics, School of Physics, Huazhong University of Science and Technology, Wuhan, 430074, China}

\begin{abstract}
   Quantum measurement is a process that involves the interaction between a quantum system and a macroscopic measurement apparatus containing many degrees of freedom. The photodetector is such an apparatus with many electrons interacting with the incoming quantum photon. Therefore the incoming photon will spread and get scrambled in the photodetector, that is, the operator of the initial incoming local photons will grow and becomes highly non-local through the interaction process. Investigating this scrambling process in detail is useful for understanding the interaction between the quantum system and the measurement apparatus. In this paper, we study the quantum scrambling process in an effective toy model of photodetectors in three different physical scenarios, by numerically simulating the evolution of the out-of-time correlators\,(OTOC). In particular, the integrability of the effective model is explored through level spacing statistics, and the effect of the spatially/temporarily distributed disorders on the system evolution is carefully investigated using the OTOC. Looking into these detailed dynamical processes paves the way to the quantum simulation and manipulation of photondetector, which would provide insights into understanding the wave-function collapse processes.
\end{abstract}
\maketitle
\section{Introduction}\label{sec:1}
The evolution of physics in the early 20th century fundamentally changed human understanding of nature, of which quantum mechanics profoundly changes our epistemology of physical reality. However, one cloud that shades the sunshine of quantum mechanics is the measurement process, during which the wavefunction of a quantum system ``collapses" when the system interacts with the apparatus\,\cite{Schlosshauer2005,Wheeler1984,Wigner1963,Bohr1928,Zurek2003,Landau1958}. This wavefunction collapse process is at the heart of the probabilistic nature of quantum mechanics. It is currently treated as an axiom of quantum mechanics which can not be derived from the Schroedinger equation. Various interpretations are proposed, but none of them is completely conclusive. For example, Copenhagen's interpretation introduces a cut on the quantum world and classical world to explain the wavefunction collapse\,\cite{Bohr1928,Heisenberg1930,Bohm1952a}, without a clear definition of the quantum/classical boundary. Hidden-variable interpretation, such as De Broglie-Bohm theory\,\cite{Bohm1952b,Bohm1952c,Bohm1963}, takes a different worldview\,\footnote{De Broglie-Bohm theory is a nonlocal hidden-variable theory, therefore does not contradict to the violation of Bell's inequality.}. It attributes the probabilistic nature of quantum phenomenon to the sub-quantum hidden variables, which obey a deterministic law. In the De Broglie-Bohm theory, the uncertainty of the measurement result comes from the complicated interaction via Bohm's ``quantum potential" between the guiding wave of the quantum system and the macroscopic apparatus which consists of many degrees of freedom\,\cite{Bohm1952b,Bohm1952c}. Everett's relative-state formulation (often referred to as the ``many-world interpretation'') does not have collapses, but simply keeps all alternative possibilities in the quantum state of the universe 
\cite{Everett1957,Dewitt1970,DeWitt1973}, while such a branching is very counter-intuitive although it is a consistent theory. Another possibility (maybe conceptually the simplest possibility) is a ``full-quantum worldview". It means that (1) the world (system and apparatus) is completely quantum-mechanical, and the world's state evolution can be completely described by the unitary evolution; (2) there is no mysterious hidden variable or world-branching; (3) the wave function collapse is a phenomenon due to the complicated interaction between the system and the apparatus. This work devotes to the exploration of this ``full-quantum worldview".

\begin{figure}[h]
\centering
\includegraphics[width=0.5\textwidth]{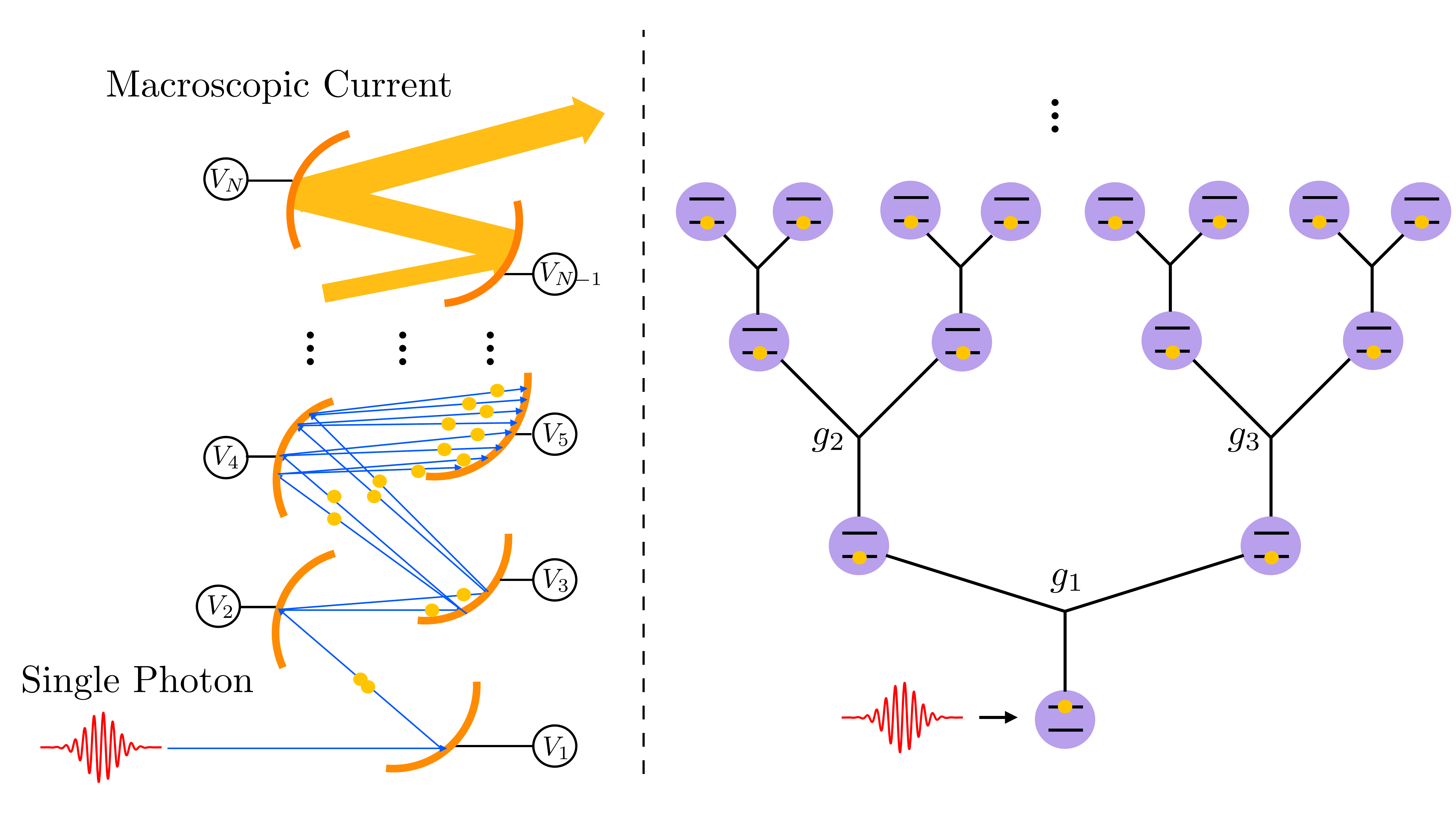}
\caption{A toy model of a photo-detector or a photo-multiplier tube. The left panel schematically shows the photon-electric avalanche process, where the electron triggered by a single photon at the first dynode will be amplified to be a macroscopic photon-electric current. The right panel is an effective toy model of the photon-detection process.}\label{fig:scheme}
\end{figure}

One unavoidable question in all quantum measurement interpretation is the interaction between the to-be-measured system and the apparatus, of which the latter contains a large number of degrees of freedom\,(a ``macroscopic system")\,\cite{Bohr1928,Landau1958}. A typical example of a quantum measurement system is the photo-detector or photo-multiplier tube\,\cite{Feynman1988}, see the left panel of Fig.\,\ref{fig:scheme}. A single photon knocks out an electron in the first photo-sensitive electrode (dynode), which is accelerated to the second dynode due to the biased electric field between these two dynodes. The accelerated electron will hit the second dynode and generate more electrons (e.g. two electrons), which will be accelerated to the third dynode, and so on. This avalanche process generates a macroscopic current, which corresponds to one of the eigenstates of the initial photon. 

In an accompanying paper\,\cite{Hu2022}, an effective toy model that manifests this avalanche process is discussed and shown in Fig.\,\ref{fig:scheme}. The work presented here is its continuation that analyse the evolution of quantum information in this effective model. In this model, the translational degrees of freedom of the electrons in the photo-detector are mimicked by discrete two-level qubits, where the electrons knocked out from different dynode are represented by the qubits at different layers. These layers are coupled via a Hamiltonian as:
\begin{equation}\label{eq:hamiltonian}
\begin{split}
\hat H=&\sum_i\hbar \frac{\omega_i}{2}\hat \sigma^{i}_Z+\hbar g_1\hat \sigma_-^{A}\hat\sigma_+^{B_1}\hat\sigma_+^{B_2} \\
&+ \hbar g_2\hat\sigma_-^{B_1}\hat\sigma_+^{C_1}\hat\sigma_+^{C_2}
 + \hbar g_3\hat\sigma_-^{B_2}\hat\sigma_+^{C_3}\hat\sigma_+^{C_4} + \cdots + {\rm h.c.}\,,
\end{split}
\end{equation}
where the $\omega_{i}$ is the eigenfrequency of the $i$-th qubit, which fluctuates for a real detector. The $g$-coefficients are the three-qubit interaction strength, which is decided by the microscopic interaction details of the electrons and dynodes thereby being also random. The $\hat\sigma_\pm=(\hat\sigma_x\pm i\hat\sigma_y)/2$ are the raising and lower operators of the qubits. It is important to note that this is a Hamiltonian containing randomly fluctuating parameters\, (details are presented in the next section).  Our accompanying paper discussed the simulation of the above model using near-term quantum computers\,\cite{Hu2022}.

In this setup, the collapse of quantum state takes place in a simple form: an exponentially decaying wavefunction gets collapsed into a pulse at a particular moment.  
\begin{equation}
|\gamma\rangle = \int dt e^{-\lambda t}|t\rangle  \rightarrow  |t_0\rangle 
\end{equation}
with probability density $p(t_0) \propto e^{-2\lambda t_0}$. The eigenstate $|t_0\rangle $ corresponds to the avalanche taking place at time $t_0$. In this paper, we will not exactly recover the behavior, but will provide arguments and preliminary numerical evidence that such evolution will take place.  In our case, it is the particular realizations for the $g_j$'s that will control when the collapse will take place. Our dynamics does not solve the collapse problem, before further issues will be addressed.  We will outline those issues at the end of this paper.


Before returning to the issue of quantum-state collapse, in the main part of this paper, we focus on the dynamics of our photodetector toy model. This toy model, which captures the avalanche property of the photodetector, is a many-body system with random parameters. Therefore it is interesting to investigate how the quantum information carried by the initial photon spreads into the measurement apparatus and gets scrambled in this network, which will be carefully discussed and simulated in the later sections. The so-called out-of-time correlator\,(OTOC) and Holevo information are useful tools to study the scrambling of the quantum information carried by the initial photon in a many-body system\,\cite{Larkin1969,Swingle2018,Braumuller2022}. In this work, we will simulate the evolution of OTOC and also the Holevo information of our model. Moreover, we also explore the integrability of our toy model by analysing the eigenvalue level spacing statistics, and the results strongly indicate that our system is integrable, even though there is an avalanching behaviour.

The structure of this paper is the following. Section\,\ref{sec:2} discusses the toy model, its Hilbert space structures and the primary picture of quantum information scrambling in this system. In Section\,\ref{sec:3}, the integrability of the model is studied using level spacing statistics. Section\,\ref{sec:4} devotes to studying the evolution of the out-of-time correlators in different environmental settings. Section\,\ref{sec:5} investigates the quantum information scrambling by simulating the Holevo information. Section\,\ref{sec:6} concludes the paper. 

\section{Model Hamiltonian and the structure of network Hilbert space}\label{sec:2}
\subsection{Model descriptions}\label{sec:2.1}
We have briefly introduced our model Hamiltonian in Section\,\ref{sec:1}. Here we give a more detailed discussion and comments on this model. The main features of our toy model of a quantum measurement process using a photo-multiplier tube can be summarised as follows. 

(1) The avalanching process in our model spreads the quantum information from a small corner of the Hilbert space to a much larger portion of the Hilbert space (with a larger number of excitation) which means the process is practically irreversible (or equivalently, the reversing time is very long). This reminds us of the Liouville flow of a classical complex system that diffuses in the phase space.

(2) Our toy model is a unitary system. This unitary setup not only fits the ``full-quantum worldview'', but also matches the real photo-detection process shown in Fig.\,\ref{fig:scheme}. The basic physical processes in Fig.\,\ref{fig:scheme} contain the acceleration of the electron and the collision of the accelerated electron with the dynode. (i) The electron acceleration happens during a very short time scale $\tau_{\rm fly}\sim l_D/v_e\sim\sqrt{m l_D^2/eU_D}\sim 20\,{\rm ps}$\,(for the inter-spacing of dynodes $l_D\sim 0.1\,{\rm mm}$, and the voltage between neighbouring dynodes to be $U_D=100$\,V, a real photo-multiplier has a nanosecond rise-up time\,\cite{Hamamatsu2007}) and it is reasonable to assume that the space between different dynodes is an ideal vacuum. This means that the electron acceleration process is unitary. (ii) The collision of the accelerated electrons with the dynode also happens in a very short time interval $\tau_{\rm col}\sim \hbar/E_{e\rm in}\sim 10^{-17}\,{\rm s}$ for a 100\,eV incoming electron as an example. During this interaction process, the energies of the electrons in the dynode can have fluctuations, which manifest as the fluctuation of the phase matching in Eq.\,\ref{eq:hamiltonian}. For example, in the interaction picture:
\be\label{eq:Hamiltonian2}
\hbar g_2\hat\sigma_-^{B_1}\hat\sigma_+^{C_1}\hat\sigma_+^{C_2}\rightarrow \hbar g_2\hat\sigma_-^{B_1}\hat\sigma_+^{C_1}\hat\sigma_+^{C_2}e^{-i(\Delta_{B1}-\Delta_{C1}-\Delta_{C2})t},
\ee
where $\Delta_{B1}-\Delta_{C1}-\Delta_{C2}\equiv \delta\Delta_{g_2}$ is fluctuating due to the environmental effects in the dynode. However, since the interaction time is very short, it is a good approximation to think that the quantum information carried by the electrons can hardly spread into the dynode environment \,(this time scale can be estimated as the electric charge equilibrium relaxation time scale $\sim 10^{-15}$\,s, which is orders of magnitude longer than the $\tau_{\rm col}$).   Therefore the phase-matching fluctuation can be treated as a parameter fluctuation in the Hamiltonian while the system is still unitary. In our model, we assume the frequency detuning $\delta\Delta_{gi}$ at each interaction vertex has a random fluctuation, and satisfies a homogenous distribution within $[-\sigma_\Delta,+\sigma_\Delta]$, in which the $\sigma_\Delta$ is denoted as the disorder strength.

(3) Besides the dynode environmental effect, our toy model also ignores the possible Coulomb interaction between the electrons during the acceleration process, which is complicated and leads to the exchange of quantum information among these electrons, especially at the late stage of the avalanche process where the electron density becomes high. Since the electron flying time $\tau_{\rm fly}$ is very short, we assume that this Coulomb interaction effect can be ignored.

(4) For a practical quantum measurement process, the macroscopic current at the later avalanche stage will be recorded by the observer. Taking into account this recording process means that this toy model must be treated in an open system, which is beyond the current scope of this work. Here we only focus on the unitary evolution of our avalanching network.

\subsection{Relationship with other models}
\begin{figure}[h]
\centering
\includegraphics[width=0.48\textwidth]{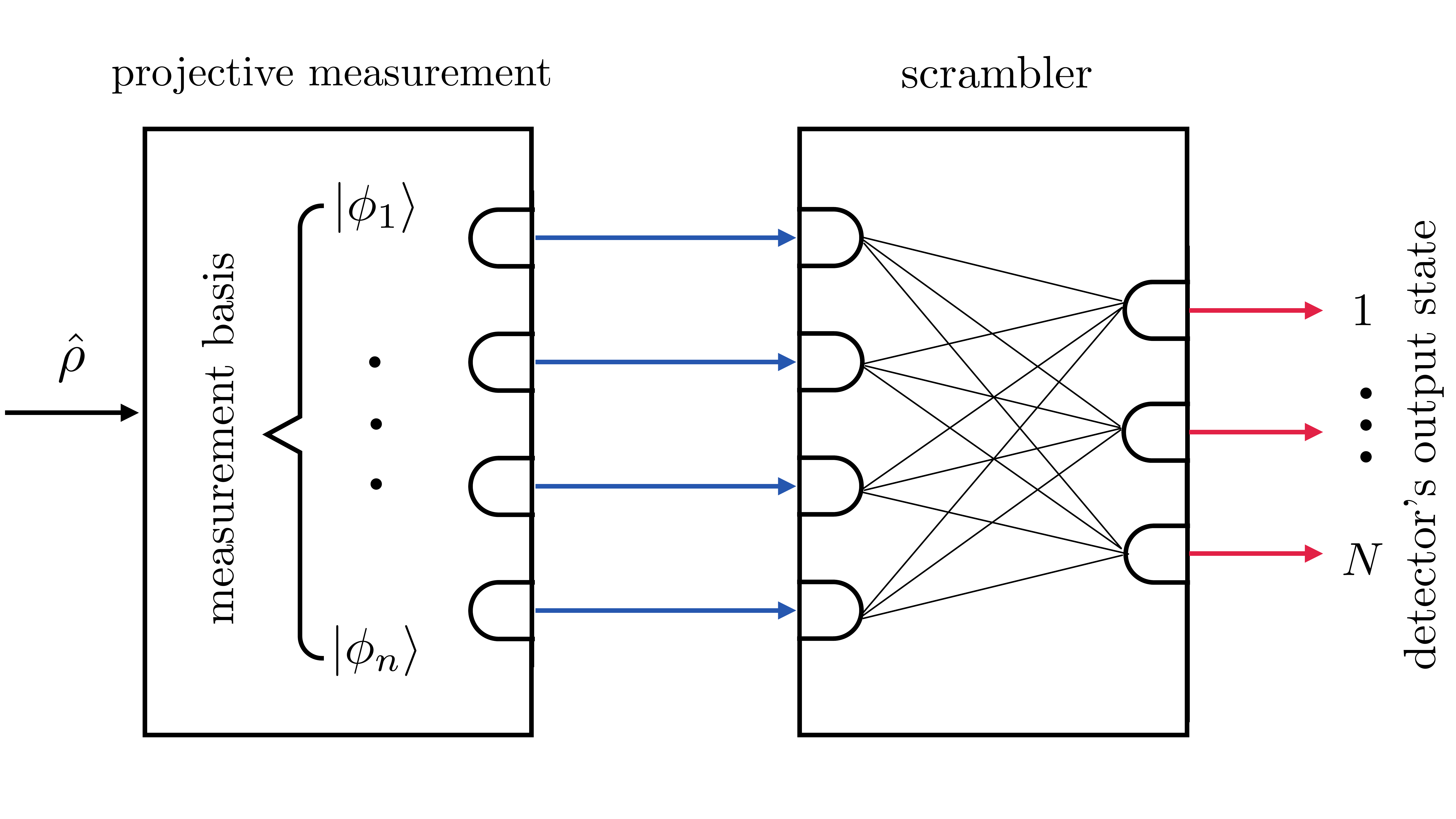}
\caption{Lvovsky's understanding of a quantum measurement process, in which he pointed out the scrambling behaviour of quantum photon-detector. This figure is re-drawed from\,\cite{Lvovsky2018}.}\label{fig:lvovsky}
\end{figure}

(1) It is worth noting that it is A. I. Lvovsky who first noticed the scrambling behaviour of a photodetector\,\cite{Lvovsky2018}, the figure which we borrowed from his book is re-drawn in Fig.\,\ref{fig:lvovsky}. The difference between Lvovsky's scrambling model and our work is the following: Lvovsky's model consists of an ideal projective measurement on the basis $|\nu_i\rangle$ and then scrambler as a classical device that randomly maps the eigenstates to the detector's outputs. Therefore in Lvovsky's model, the scrambling process is entirely classical. However, the scrambling process in our model is inherent in the qubit-network unitary evolution with a random Hamiltonian. 

(2) This model is reminiscent of the famous Fermi-Pasta-Ulam-Tsingou\,(FPUT)\,\cite{Fermi1955,Dauxois2008} model in classical nonlinear dynamics, which is widely used in understanding the thermalisation problem\,\cite{Berman2005}. By studying a chain of $N$ non-linear interacting harmonic oscillators with the initial condition that the energy is stored in one of the harmonic oscillators, FPUT tried to demonstrate that the irreversibility of a thermalisation process can emerge from local reversibility. FPUT found a quasi-periodic appearance of specific states which is the so-called ``\emph{Fermi-Pasta-Ulam-Tsingou recurrence}". Our toy model is similar in the sense that the harmonic oscillators in the FPUT model are replaced by the qubits, and the FPUT recurrence will also exhibit in our model\,(see Section\,\ref{sec:3}). Note that despite the similarity of our model and the FPUT model in the interaction form,  our model is  inherently linear since quantum mechanics is linear.

(3) The topology of our model may remind the reader about the Cayley tree or Bethe lattice, which has been studied in\,\cite{Anderson1980,Shapiro1983}. Though the similarity in the configuration topology, the Hamiltonian mainly consists of near-site hopping and random chemical potential, which is different from our model since it does not have avalanche behaviour. The Anderson model was used to explain the electron conductivity in solid materials with disorders. Moreover, the randomness in our model could be time-dependent.

\subsection{Reduced network Hilbert space}
\label{sec:2.2}
Suppose we have $\mathcal{N}$ qubits in this network, and initially only the first qubit is occupied, which corresponds to the excitation of the photo-electron in the first dynode. The dimension of the Hilbert space of the entire network is $2^{\mathcal{N}}$, which means we should simulate $2^{\mathcal{N}}\times2^{\mathcal{N}}$  matrix evolution. However, our three-qubit interaction Hamiltonian has a special conservation quantity: 
\be
\bar{N}=\sum_j\sum^{N_j}_{i_j=1}\frac{\hat n_{i_j,j}}{2^{j-1}},
\ee
where $j$ is the index of different layers and the $i_j$ labels the $i$-th site of the $j_{\rm th}$ layer and $\hat n$ is the occupation operator. For example, we have $\bar N=1$ when our network is triggered by an excitation on the first qubit,
which means only a very small portion of the Hilbert space of the network will be involved. All those states that violate this conservation law will not participate in the network evolution. The dimension $D$ of the conserved subspace of the $j$-th layer satisfies the following recurrence equation:
\be
D[N_j]=D[N_j-1]^2+1.
\ee
For example, for 6 layers with 63 qubits, the interested reduced Hilbert space has a dimension of 458330, which is only a $\sim 10^{-15}$ part of the full Hilbert space. This structure greatly simplifies our numerical simulation.

\begin{figure}[h]
\centering
\includegraphics[width=0.48\textwidth]{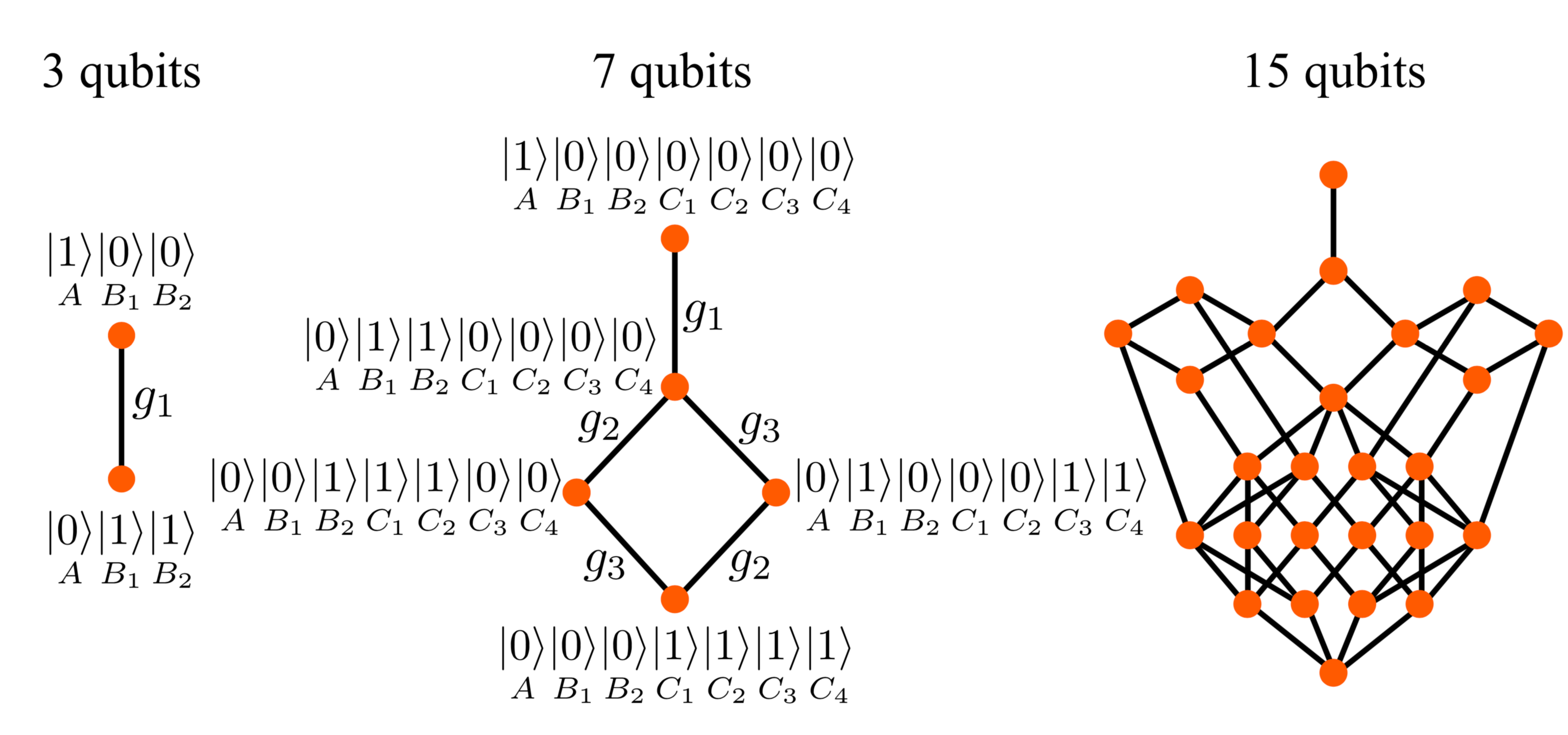}
\caption{Graphic representation of the Hamiltonian matrix in the conserved subspace for different network sizes. The red spots represent different basis states of the system and the links represent the interactions that connect these basis states.}\label{fig:reduced_hilbert}
\end{figure}

In order to study the quantum information scrambling in our toy model, we need to construct an algorithm to relate the reduced network Hilbert space and the full Hilbert space.

For presenting the general formula for relating the reduced network Hilbert space and the full network Hilbert space, some definitions are given as follows. First, we define the $\tilde{|i\rangle}_N$ as the $i_{\rm th}$ basis in the \emph{reduced Hilbert space} of a $N$-layers configuration. Then we define an operator $\hat{O}_N(i)$ acts on the zero-excitation ground state $|0\rangle^{\otimes2^{\mathcal{N}}-1}$ and generates a basis $\tilde{|i\rangle}_N$ in the reduced Hilbert space:
\begin{equation}
\tilde{|i\rangle}_N=\hat{O}_N(i)|0\rangle^{\otimes2^{\mathcal{N}}-1}
\end{equation}

\begin{figure}[h]
\centering
\includegraphics[width=0.5\textwidth]{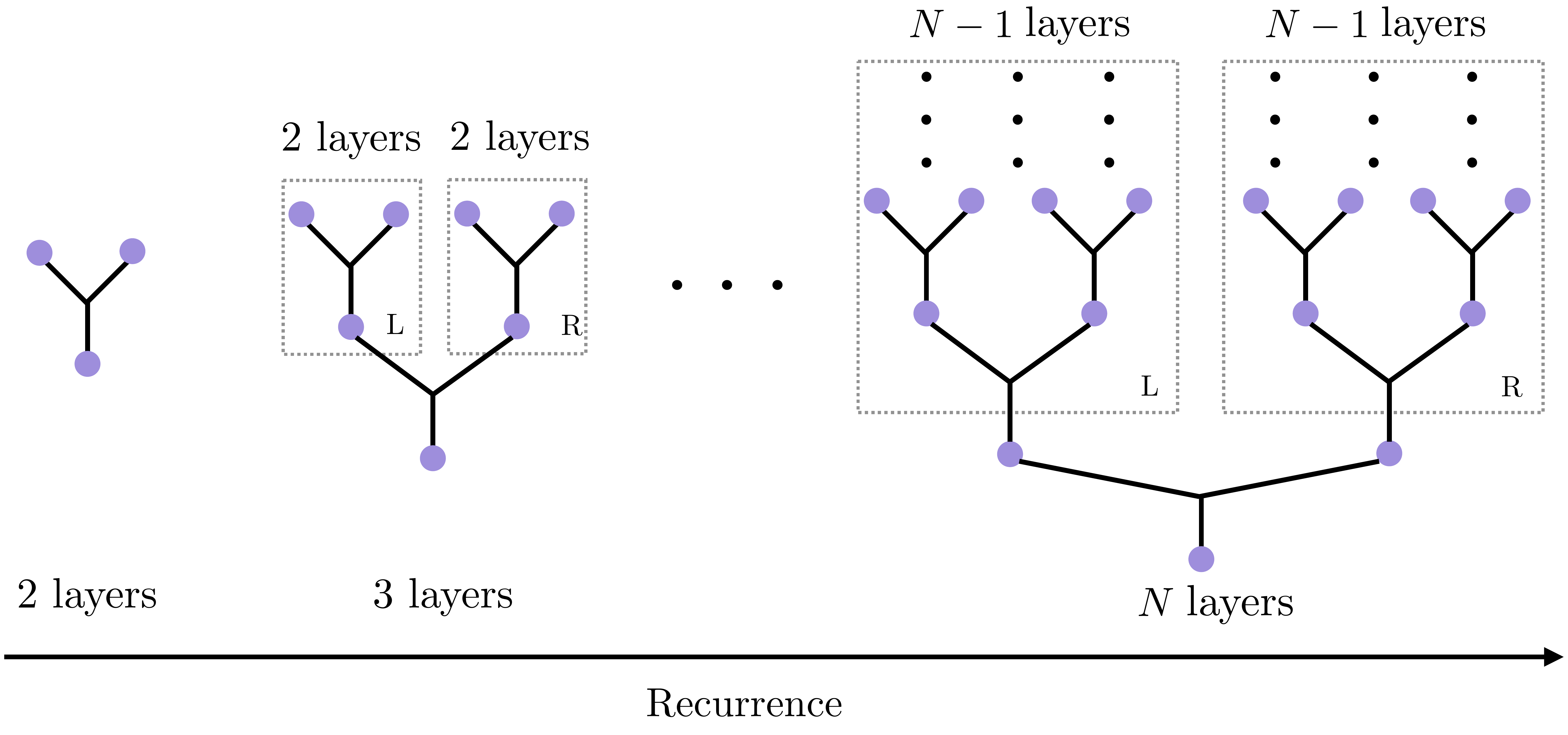}
\caption{Graphic illustration of the algorithm to relate the states in the reduced network Hilbert space and the full Hilbert space. Note that the blue spot here represent the qubit in the tree, not the basis states as in Fig.\,\ref{fig:reduced_hilbert}}\label{fig:algorithm}
\end{figure}

It is shown in Fig.\,\ref{fig:algorithm} that the $N$-layers structure can be split into two $N-1$-layers configurations with a top qubit. We can now define an operator $\hat{O}^{R/L}_{N-1}(i)$ acts on the right/left $N-1$ sub-layers of the original $N$-layers structure, and generates the state with qubit excitation in a similar way as $\hat{O}_{N-1}(i)$. Then it is crucial to find the order of the basis of the reduced Hilbert space with excitation $\hat{O}^R_{N-1}(m)$ and $\hat{O}^L_{N-1}(m)$ in the \emph{$N-1$ sub-layer structures} in the reduce Hilbert space of the \emph{$N$-layer structure}. The algorithm that fulfills this task is by using the recursive formula:
\begin{equation}
\begin{split}
&\hat{O}_N((m-1) D[N-1]+n+1)=\\
&\begin{cases}
             \hat\sigma^1_+,&m=1,n=0;\\
             \hat{O}^{R}_{N-1}(m)\hat{O}^{L}_{N-1}(n),&
              m,n\in \{1,..., D[N-1]\};
             \end{cases}
\end{split}
\end{equation}
where the $\hat\sigma^1_+$ is the flip operator of the top qubit. To understand this algorithm, we can determine the order in the $N$-layer structure as follows\,(see Fig.\,\ref{fig:algorithm}): for the $m_{\rm th}$ basis of the left $N-1$ sub-layer to step forward to the $m_{\rm th}+1$ state, one need to go through all $D[N-1]$ basis of the right sublayer. This raises the order index of the corresponding basis of the $N$-layer structure by $D[N-1]$. The iteration leads to a $(m-1)D[N-1]$ index, adding by an $n+1$ denotes the states of the right sub-layer and the top qubit.

For realising this recurrence algorithm, the $\hat{O}_2(i)$ is needed as an initial condition. In this paper, we choose $\hat{O}_2(i)$ by following way,
\begin{equation}
\begin{split}
\tilde{|1\rangle}_2&=\hat{O}_2(1)|0\rangle^{\otimes3}=|1\rangle_1|0\rangle_2|0\rangle_3,\\
\tilde{|2\rangle}_2&=\hat{O}_2(2)|0\rangle^{\otimes3}=|0\rangle_1|1\rangle_2|1\rangle_2,
\end{split}
\end{equation}
where $|1/0\rangle_i$ represents that the $i_{\rm th}$ qubit is one/no excitation.

To simulate the operator evolution later, we need to construct the operator representation in the reduced Hilbert space. Note that some local operators can not be represented in reduced Hilbert space in a closed way, because some states are out of the reduced Hilbert space.  For example, The local operator $\hat\sigma_{+}^{1}$ in two layers case, acting on the state $|0\rangle_0|1\rangle_2|1\rangle_3$, will transfer it to $|1\rangle_1|1\rangle_2|1\rangle_3$, which is out of the reduced Hilbert space. Fortunately, there exist some operators of which the representation always stays in the reduced Hilbert space, such as $\hat{\sigma}_i^z$, which will be used in constructing the out-of-time correlator later on.  
The formula of the matrix elements are given as:
\begin{equation}
\begin{split}
\tilde{\langle m|}\hat{\sigma}_i^z\tilde{|n \rangle}=\begin{cases}
        0,&m\neq n;\\
        1,&m=n, {\rm Tr}_{\bar{i}}(\tilde{|m\rangle}\tilde{\langle m|})=|1\rangle_i\langle 1|_i;\\
        -1,&m=n, {\rm Tr}_{\bar{i}}(\tilde{|m\rangle}\tilde{\langle m|})=|0\rangle_i\langle 0|_i;
        \end{cases}
\end{split}
\end{equation}
where ${\rm Tr}_{\bar{i}}(...)$ is tracing over the other qubits complementary to the $i_{\rm th}$ qubit.

\subsection{Quantum scrambling}
Quantum information of the incoming photon is encoded in the first qubit with the initial state as $|\psi\rangle=\alpha|0\rangle+\beta|1\rangle$, where the coefficients $\alpha,\beta$ carry the initial unscrambled quantum information. Quantum information scrambling means that once this qubit interacts with the other qubit in the network that mimics the photo-detector, one can not recover the $\alpha,\beta$ from any \emph{local measurement} of this qubit network. A simple illustration is that the quantum state at time $t$ can have the structure as:
\be
\begin{split}
&|\psi(t)\rangle=\alpha|0\rangle_1\otimes\prod_{ij}|0_{ij}\rangle+\beta|\psi\rangle_{\rm scr},\quad{\rm with}\\
&|\psi\rangle_{\rm scr}=c_0|1\rangle\otimes\prod_{ij}|0_{ij}\rangle+...+\sum_m c_m|0\rangle\otimes|...1_{4j}...\rangle_m\\
&\qquad\qquad+\sum_n c_n|0\rangle\otimes|...0_{4j}...\rangle_n+...,
\end{split}
\ee
where the $|\psi\rangle_{\rm scr}$ represents the state with the avalanching process, and $|...1/0_{4j}...\rangle$ represents the state where the $j_{\rm th}$ site of the 4$_{\rm th}$ layer is occupied/unoccupied.

Measuring the local site $4j$ leads to the probability of occupied and unoccupied $4j$-site as $P_{1}=|\beta|^2\sum_m|c_m|^2$ and $P_0=|\alpha|^2+\beta \sum_n|c_n|^2$. We also have $|\alpha|^2+|\beta|^2=1$ and the normalisation condition for the coefficients in $|\psi\rangle_{\rm scr}$. These equations are not sufficient to solve the $|\alpha|^2,|\beta|^2$, which demonstrates that local measurement can not recover the initial quantum information and therefore the detector network scrambles the quantum information.

Taking into account the time-dependent randomness of the qubit level, after a sufficiently long evolution time, as we shall see in the later simulations, the $|\psi\rangle_{\rm scr}$ evolves into a heavily scrambled state\,(note that the corresponding density matrix is proportional to a unitary matrix), but still unitary. Therefore the total quantum state is:
\be
|\psi(t)\rangle=\alpha|0\rangle_1\otimes\prod_{ij}|0_{ij}\rangle+\beta|\psi(t\rightarrow\infty)\rangle_{\rm scr},
\ee
which is a scrambled macroscopic Schroedinger cat state. In the next section, we will use the out-of-time correlator to quantitatively study the quantum scrambling in our toy model.

\section{Level Statistics and integrability}\label{sec:3}
\begin{figure}
\centering
\includegraphics[width=0.45\textwidth]{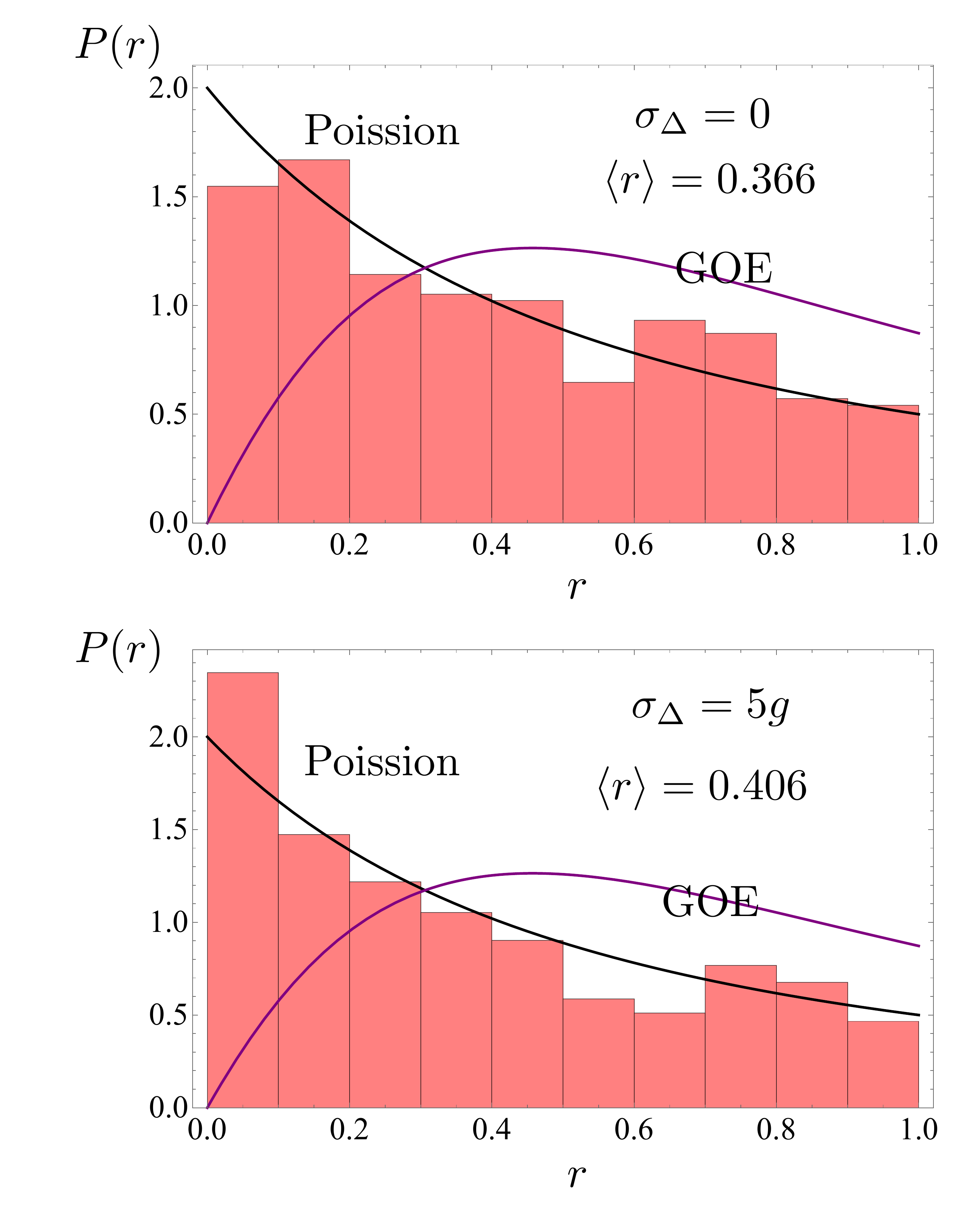}
\caption{Level spacing in our model with five layers (667 basis-states in total). Upper panel: the ideal case when the phase matching in each interaction kernel is perfect. Lower panel: the phase matching takes a random number distributed among interaction kernels with $\sigma_\Delta=5g$. It is clear that the level spacing in our system follows a Poisson distribution which indicates that our system is integrable.}\label{fig:level_spacing}
\end{figure}

Before introducing our simulation result, it is worth investigating the integrability of our model. In the structure of the reduced Hilbert space (corresponds to the conserved quantity $\bar N=1$) shown in Fig.\,\ref{fig:reduced_hilbert}, the link between different red spots physically represents a non-zero matrix element that connect these basis-states due to the interaction kernels. Therefore in the reduced Hilbert space, our model (triggered by one excitation on the first layer) can be rewritten as:
\be
\hat H_{\rm reduced}=\sum_i E_{\tilde i}|\tilde i\rangle\langle \tilde i|+\sum_{\langle i,j\rangle}g_{ij}|\tilde i\rangle\langle \tilde j|,
\ee
where $\langle i,j\rangle$ means summation over all allowed connections in the graph of Fig.\,\ref{fig:reduced_hilbert}. This bilinear Hamiltonian is diagonalisable, therefore one can obtain its eigenvalue and level spacing statistics defined as\,\cite{Pal2010,Oganesyan2007}:
\be
r_{n}=\frac{{\rm min}[\delta_{n},\delta_{n+1}]}{{\rm max}[\delta_{n},\delta_{n+1}]},
\ee
where $\delta_{n}=|E_n-E_{n-1}|$ is the energy difference of the adjacent levels. We compute the probability distribution of $r_n$ using a five-layer model with $667$ basis-states for the ideal case and the case with random phase distribution, obtaining Poisson distributions for both cases shown in Fig.\,\ref{fig:level_spacing}. For the model with more layers, e.g.\,6 layers, the number of basis would increase to $458330$, and our current computational resources can not diagonalise a matrix of this size. However, our result on the 5-layer system still indicates strongly that our system is integrable. Therefore, although our system has a avalanching beahviour that involves an exponentially growing number of operators, this Poission distribution means that our system seems to be not chaotic.

\section{Simulation Results: Evolution of the toy model}\label{sec:4}

\emph{Out-of-time correlator}\,(OTOC) is initially introduced by Larkin and Ovchinnikov in the 1960s where they used the quasi-classical method to study the electron behaviour in a superconductor with impurities\,\cite{Larkin1969}. Its importance was widely appreciated in recent years when its relevance to chaos in the quantum complex system was revealed\,\cite{Swingle2018,Hashimoto2017,Braumuller2022,Maldacena2016}. The OTOC $F_{ij}(t)$ in a quantum complex system is defined as:
\be
F_{ij}(t)=\langle \psi|\hat W_i^\dag(t)\hat V_j^\dag\hat W_i(t)\hat V_j|\psi\rangle,
\ee
where $\ket{\psi}$ is the initial state of the system at $t=0$, $\hat W_i,\hat V_j$ are two operators acting on the site $i,j$ at $t=0$, respectively. The $\hat W_i(t)$ is the time-evolving operators defined as $\hat W_i(t)=e^{i\hat H t}\hat W_i e^{-i\hat H t}$. The physical meaning of OTOC can be understood from its connection to the operator commutation square:
\be
\begin{split}
C_{ij}(t)&=\bra{\psi}[\hat W_i(t),\hat V_j]^\dag[\hat W_i(t),\hat V_j]\ket{\psi}\\
&=2(1-{\rm Re}[F_{ij}(t)]).
\end{split}
\ee
where the OTOC is real for Hermitian operators $\hat W_i,\hat V_j$. Scrambling in a quantum complex system can be understood as the growth of the initial ``simple'' operators in terms of size and complexity. This operator growth means the initial operator correlates with more and more operators in the system, and it is in this way that the quantum information spreads in the system. For example, initially commuting operators $[\hat W_i,\hat V_j]=0$ becomes non-commuting when $\hat W_i\rightarrow e^{i\hat H t}\hat W_i e^{-i\hat H t}$, that makes the $C_{ij}(t)\neq 0$ and ${\rm Re}[F_{ij}(t)]<1$. The operator correlation diminishes when $C_{ij}(t)\rightarrow 0$, or $F_{ij}(t)\rightarrow 1$.

For our toy model, the Pauli operators at different sites $\hat\sigma^z_i$ is a natural choice for the $\hat W$ and $\hat V$ operators, because the dynamical process in this toy model mainly consists of the excitation and de-excitation of the qubits. In particular, we choose $\hat V=\hat \sigma^z_1(t=0)$ which is the Pauli-$z$ operator for the first qubit and describes the electron knocked out from the first dynode by the incoming photon, and $\hat W(t)=\hat \sigma^z_i(t)$ which is the Pauli-$z$ operator of the qubit in other layers.

Besides the OTOC, the mean excitation $\langle\psi|(1+\hat\sigma^z_i)/2|\psi\rangle$ can also be used to describe the evolution of the toy model. In the following, we numerically plot the dynamical evolution of the OTOC and the mean excitation of our toy model in Fig.\,\ref{fig:scheme}.  We try to understand the behaviour of our model in three different physical scenarios: (1) The ideal case where no randomness exists in our model. (2) There will be time-independent randomness of the qubit energy levels in our model. and (3) There will be time-dependent randomness of the qubit energy levels.  It is important to note that both the (1) and (2) scenarios in later sections\,\ref{sec:4.1} and \ref{sec:4.2} in principle should satisfy the quantum Poincare recurrence theorem\,\cite{Percival1961,Bocchieri1957,Schulman1978}\,(note that this should be distinguished from the FPUT recurrence\,\cite{Berman2005}). Poincare recurrence theorem states that for a time-independent quantum system with discrete energy eigenstates, after a sufficiently long but finite time, return to a quantum state that is exactly equal to the initial quantum state\,\cite{Zermelo1896,Rauer2018,Essler2016}. However, this condition would be broken if the system Hamiltonian becomes time-dependent, as we will show in the next paragraph. In this case, the quantum Poincare recurrence will be broken and the system undergoes an ergodic evolution, which leads to the heavily scrambled Schroedinger cat state.

\subsection{Fermi-Pasta-Ulam-Tsingou recurrence}\label{sec:4.1}
\begin{figure}
\centering
\includegraphics[width=0.48\textwidth]{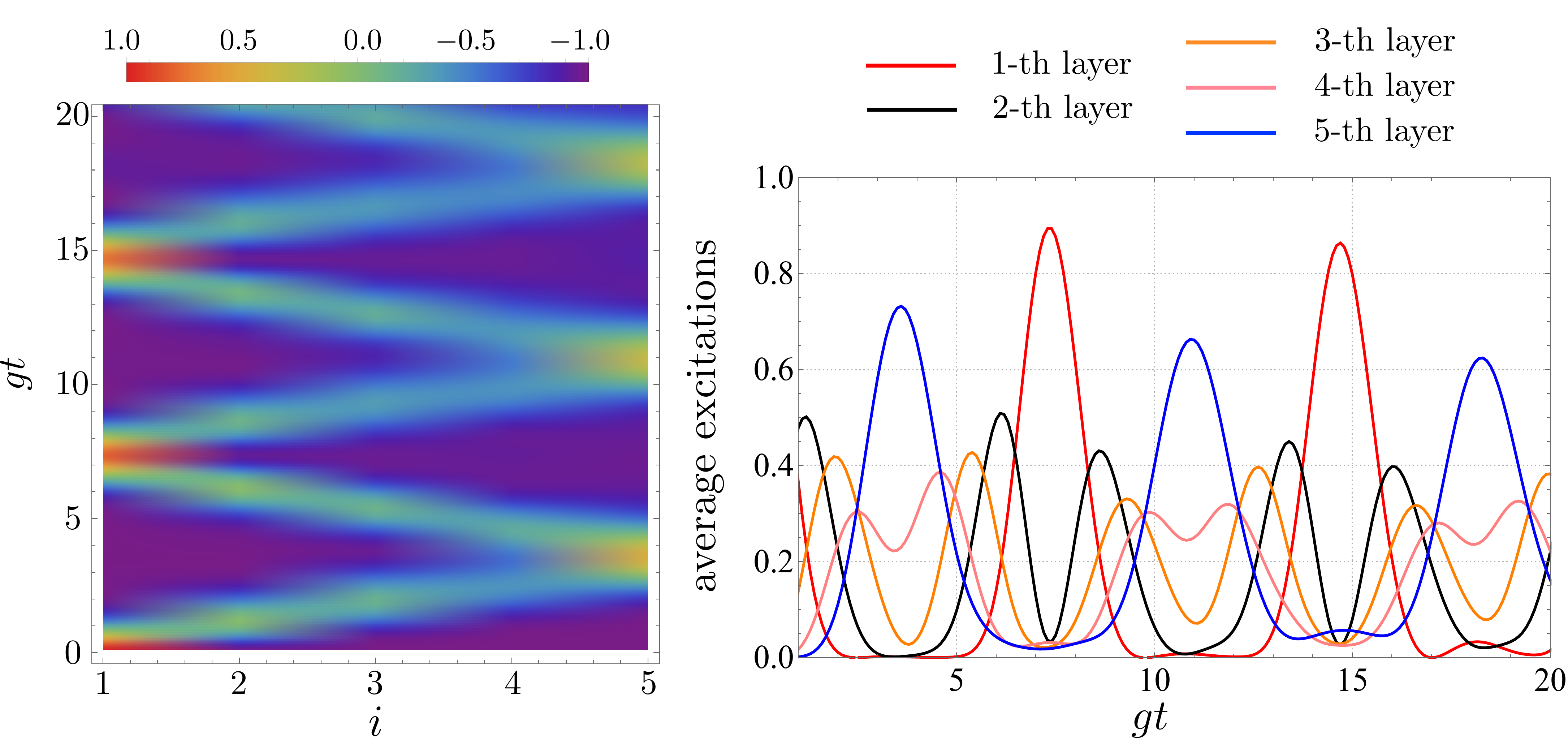}
\includegraphics[width=0.4\textwidth]{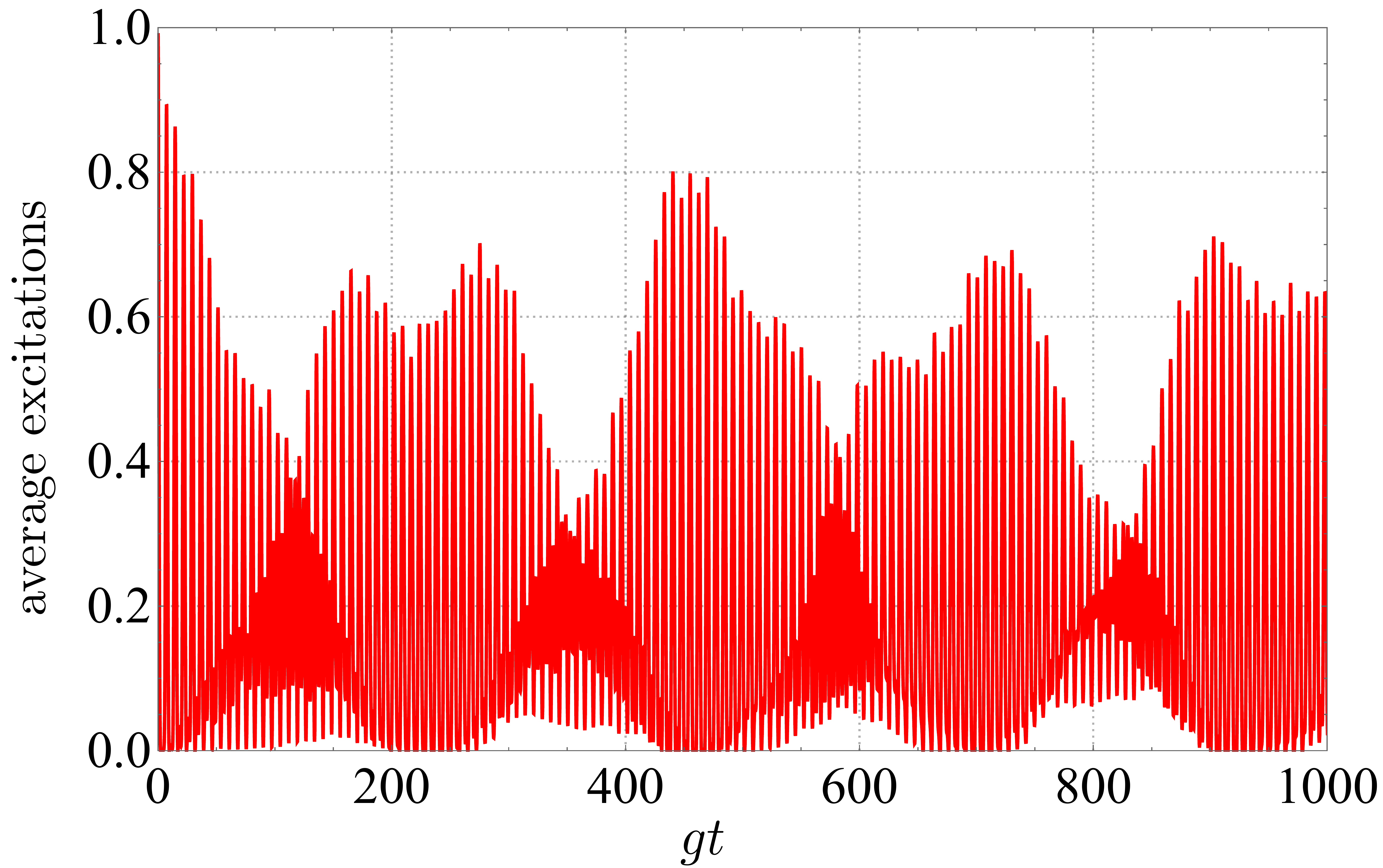}
\caption{FPUT recurrence of the ideal model without qubit level fluctuations. The upper panel is the space-time evolution of the mean occupation number of each layer simulated for a different time duration, where the ballistic propagation of the excitation and its reflection on the boundary are demonstrated. The lower panel is the evolution of the mean occupation of the first layer for a longer time scale. The features of quasi-periodicity and the super-recurrence are very similar to what happens in the classical Fermi-Pasta-Ulam-Tsingou model. The corresponding OTOC has the same behaviour.}\label{fig:FPUT_recurrence}
\end{figure} 

We first consider the ideal case described by the Hamiltonian Eq.\,\eqref{eq:hamiltonian} with a perfect environment, which means the energy level of the qubit in the dynode does not fluctuate hence the additional phase in Eq.\,\eqref{eq:Hamiltonian2} is zero. Initially, the qubit in the first layer is excited by an incoming photon. Then the mean excitation and the OTOC induced by the excitation spreads \emph{ballistically} over the system with velocity $v\propto g$,  which will be affected by the disorders introduced later.  Larger interaction strength means a quicker propagation of the quantum information into the layers. When the excitation/OTOC propagates to the boundary of the system, the flow of quantum information will \emph{quasi}-periodically bounce back and forth later on without information scrambling as shown in Fig.\,\ref{fig:FPUT_recurrence}, which is similar to the FPUT recurrence in classical dynamics\,\cite{Fermi1955}. In Fig.\,\ref{fig:FPUT_recurrence}, the evolution of the averaged excitation is shown for both short and long time scales. The results exhibit a clear FPUT recurrence structure in a relatively short time scale and super-recurrence structure in a longer time scale\,\cite{Tuck1972,Pace_2019}. 

\subsection{``Localisation" of quantum information.} \label{sec:4.2}

\begin{figure}
\centering
\includegraphics[width=0.5\textwidth]{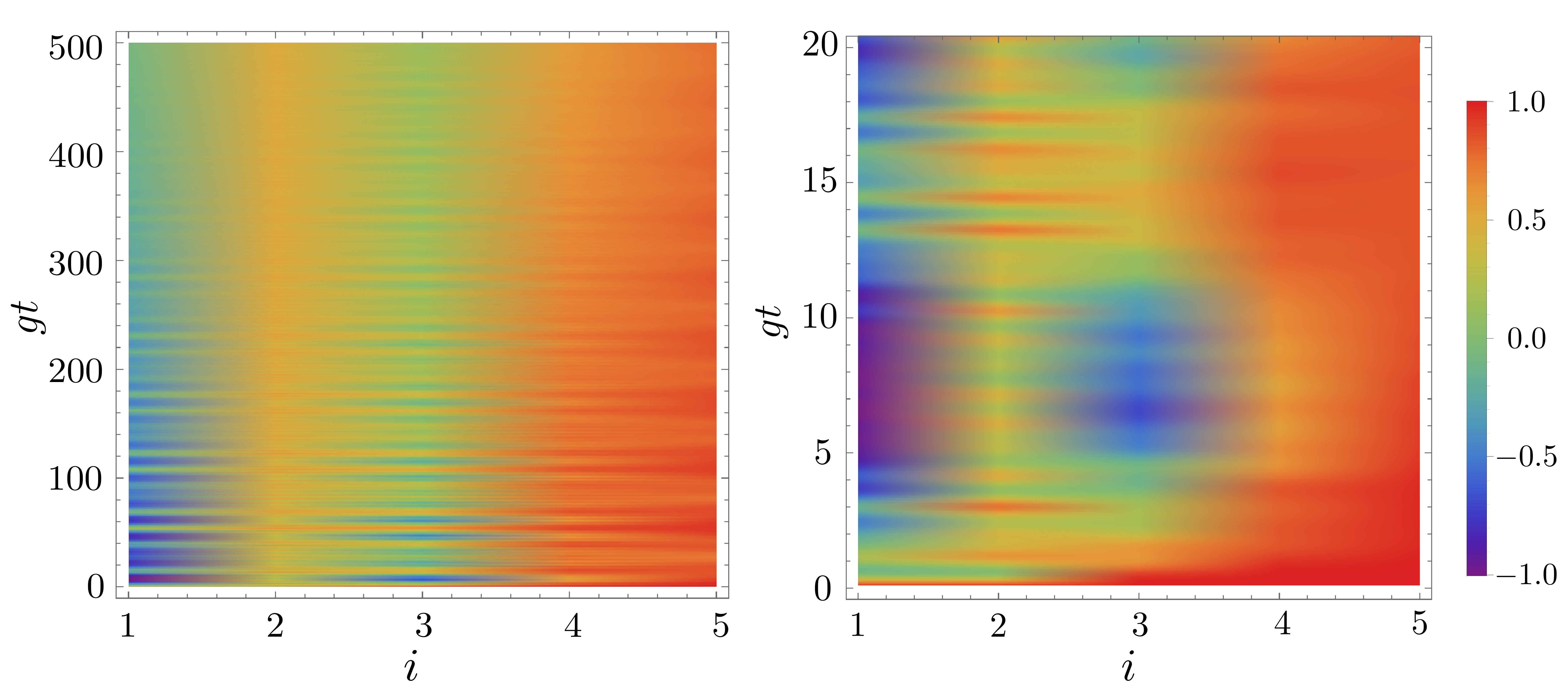}
\includegraphics[width=0.45\textwidth]{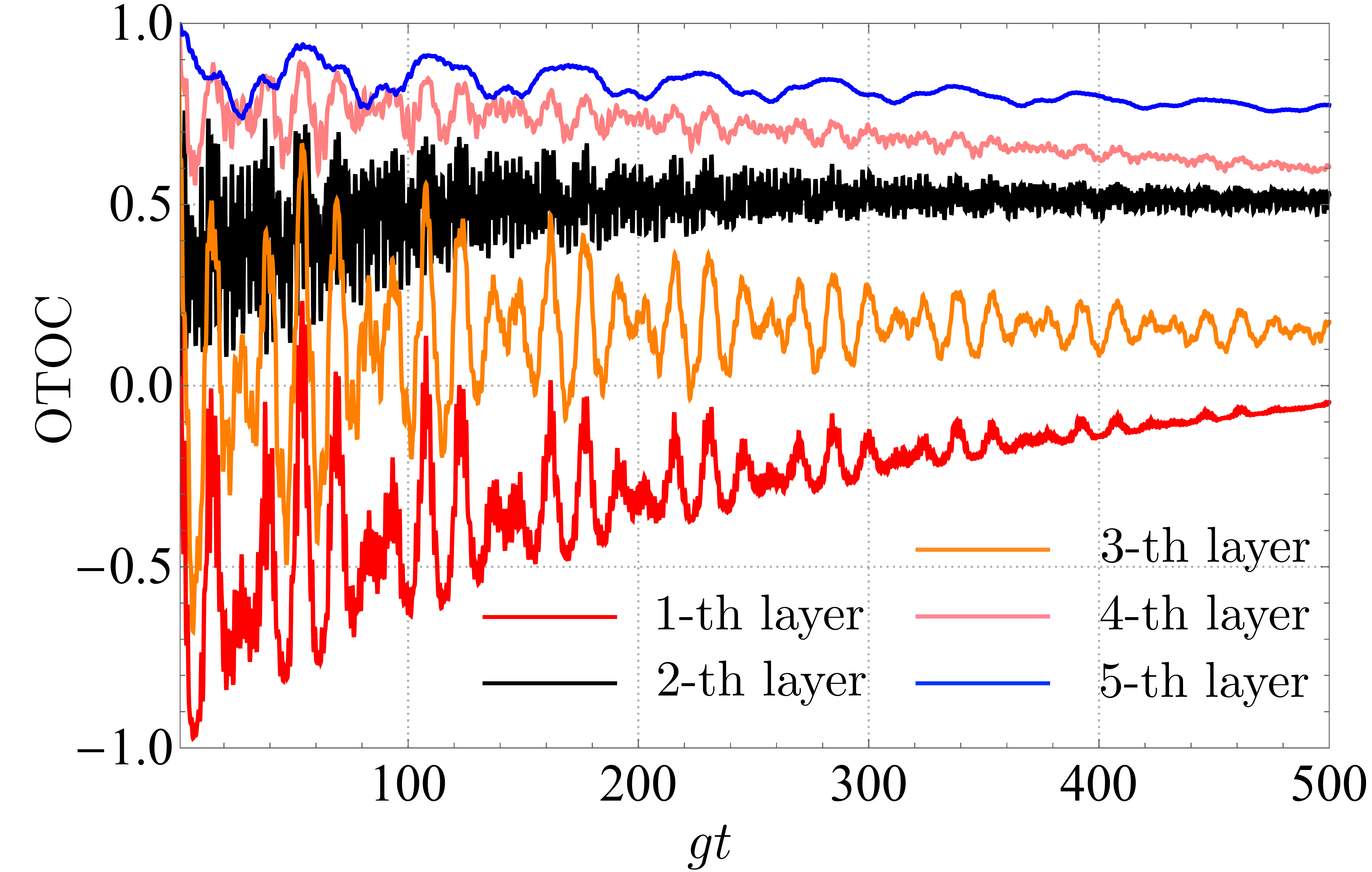}
\caption{OTOC evolution of the toy model with time-independent qubit level fluctuations among different sites, \emph{for one realisation of the disorder}. The disorder strength is chosen to be $\sigma_\Delta=5g$. The upper panel is the space-time evolution of the OTOC simulated for a different time duration, where the ``localisation" of the quantum information can be seen. The Lower panel is the OTOC evolution for different layers. It shows that the OTOC of the last two layers does not decrease to zero, and most of the quantum information is localised within the first and third layers.}\label{fig:localisation}
\end{figure}

The environmental effect is characterised by the fluctuating qubit energy levels, which will significantly affect the recurrence structure shown above. We first discuss the case when there is a time-independent energy level fluctuations randomly distributed among all qubits, with the OTOC evolution shown in Fig.\,\ref{fig:localisation} for one specific random qubit level realisation. 

The randomness of ``chemical potential"  creates a phenomenon similar to the Anderson localisation or many-body localisation discussed in the quantum thermalisation problem\,\cite{Anderson1958,Altman2018,Alet2018,Abanin2019}. The out-of-time correlator is also used in studying many-body system with randomness\,\cite{Huang2017,Swingle2017,Slagle2017,Kim2014,Fan2017,Chen2016}.
Anderson localisation is studied by imposing the Anderson model on a lattice system\,(even for the Cayley tree similar to our model), where the Hamiltonian contains a hopping term between neighbouring sites and a random qubit energy level:
\be
\hat H_{\rm Anderson}=\sum_i\hbar\frac{\omega_i}{2}\hat\sigma_z^i+\hbar g_1\hat\sigma_-^A(\hat\sigma_+^{B_1}+\hat\sigma_+^{B_2})+...,
\ee
where the mean occupation number is conserved during the interaction process for those hopping interactions. The system evolves as a wave propagating in a random medium. The randomness of the energy level will halt the spreading of the initial quantum information over different sites, which corresponds to the Anderson localisation\,\cite{Jackson2012}.

\begin{figure}[h]
\centering
\includegraphics[width=0.5\textwidth]{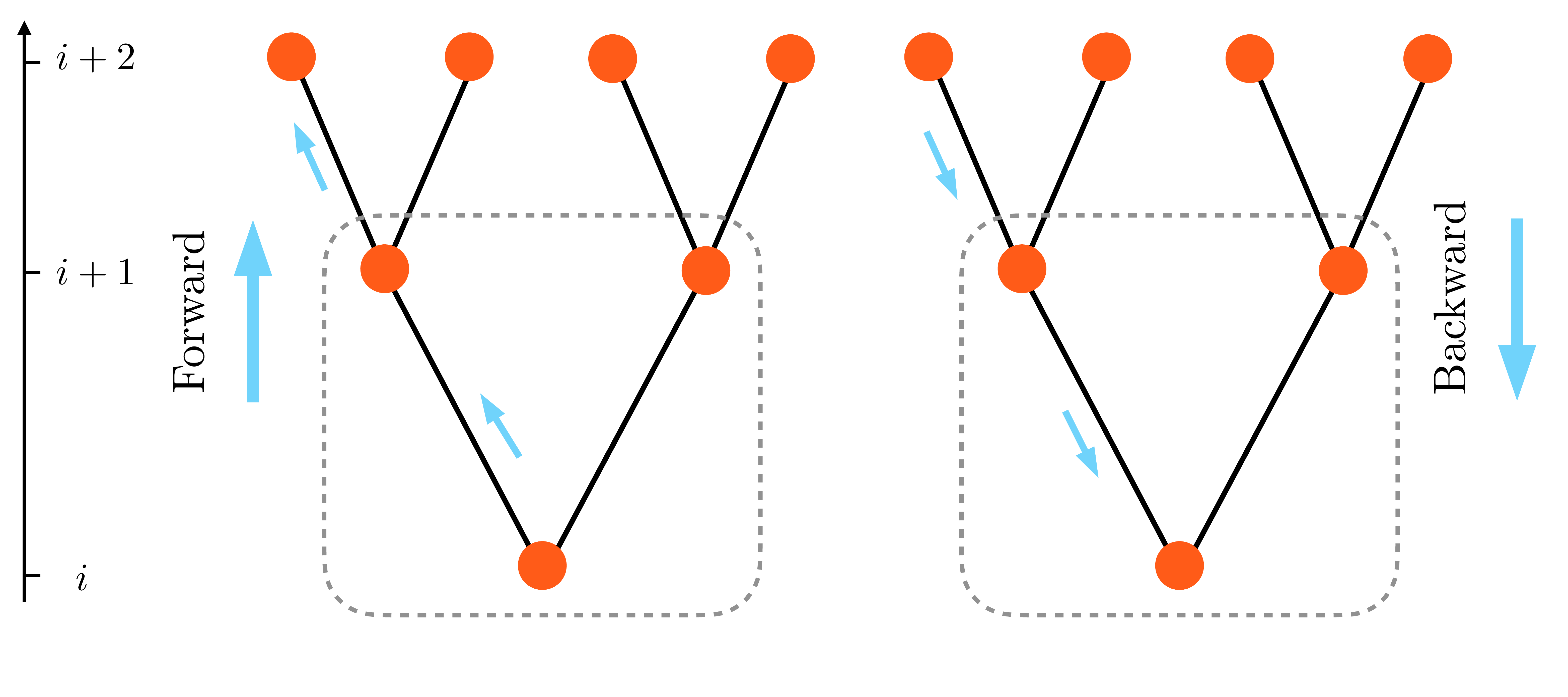}
\includegraphics[width=0.5\textwidth]{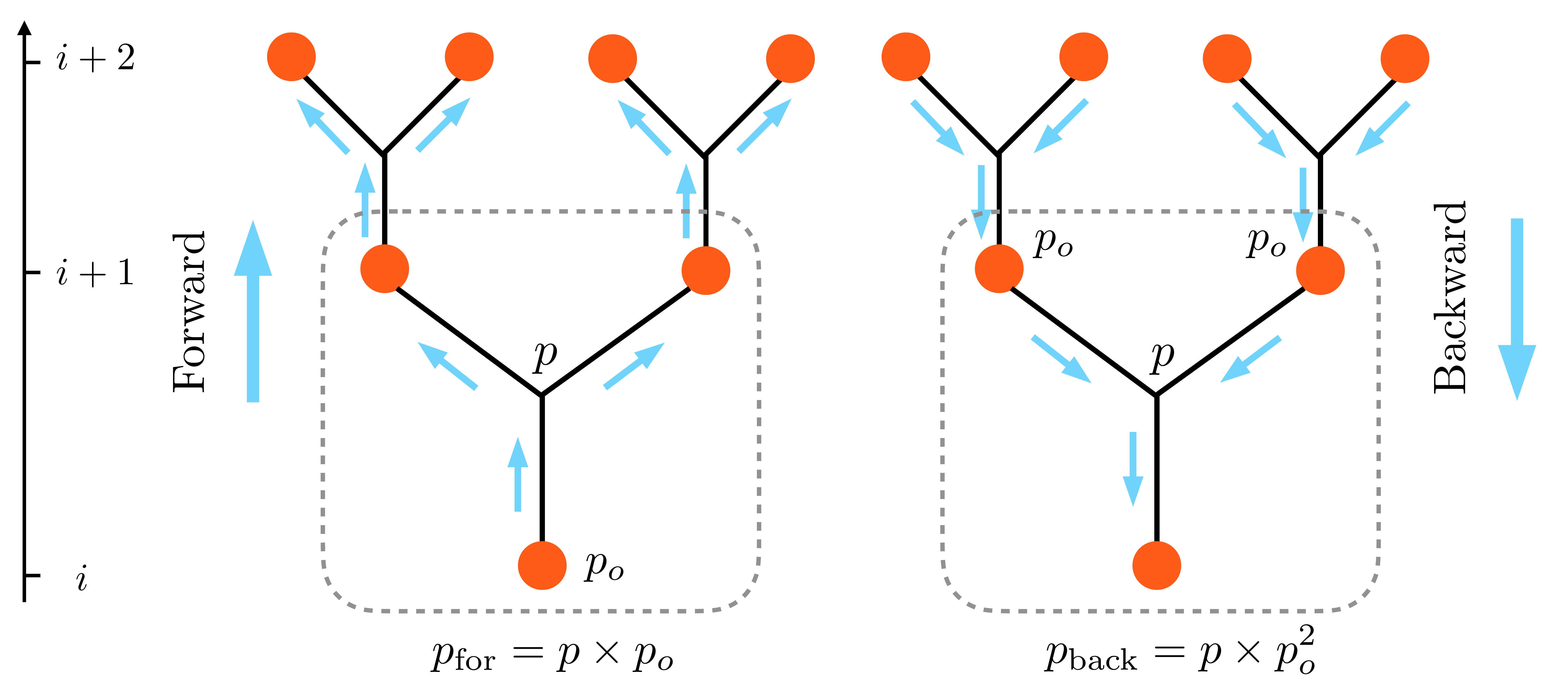}
\caption{Phase matching fluctuation induces the system irreversibility. The upper panel: the Anderson model imposed on the qubit lattice in a Cayley tree, where the probability for the phase matching of the forward process and the backward process is the same. The lower panel is on our toy model. The randomly fluctuating phase matching error favours the forward process rather than the backward process in terms of interaction efficiency.}\label{fig:irreversibility}
\end{figure}

The difference between the Anderson model Hamiltonian and our toy model brings an important difference in their evolution,  despite their similarity. Basically, the upper panel of Fig.\,\ref{fig:irreversibility} shows that,
for an occupation at the $i_{\rm th}$ layer to efficiently go to the $j_{\rm th}$ site at the $i+2$-layer, the phase matching of interactions $\propto \hat\sigma^{i,j}_-\hat\sigma^{i+1,j}_+$ and $\propto \hat\sigma^{i+1,j}_-\hat\sigma^{i+2,j}_+$ must be satisfied, with the probability $p_{\rm for}=p^2$, where $p$ is the probability for the phase matching of each interaction kernel. The probability $p_{\rm back}$ for the backward process is the same.

\begin{figure}
\centering
\includegraphics[width=0.48\textwidth]{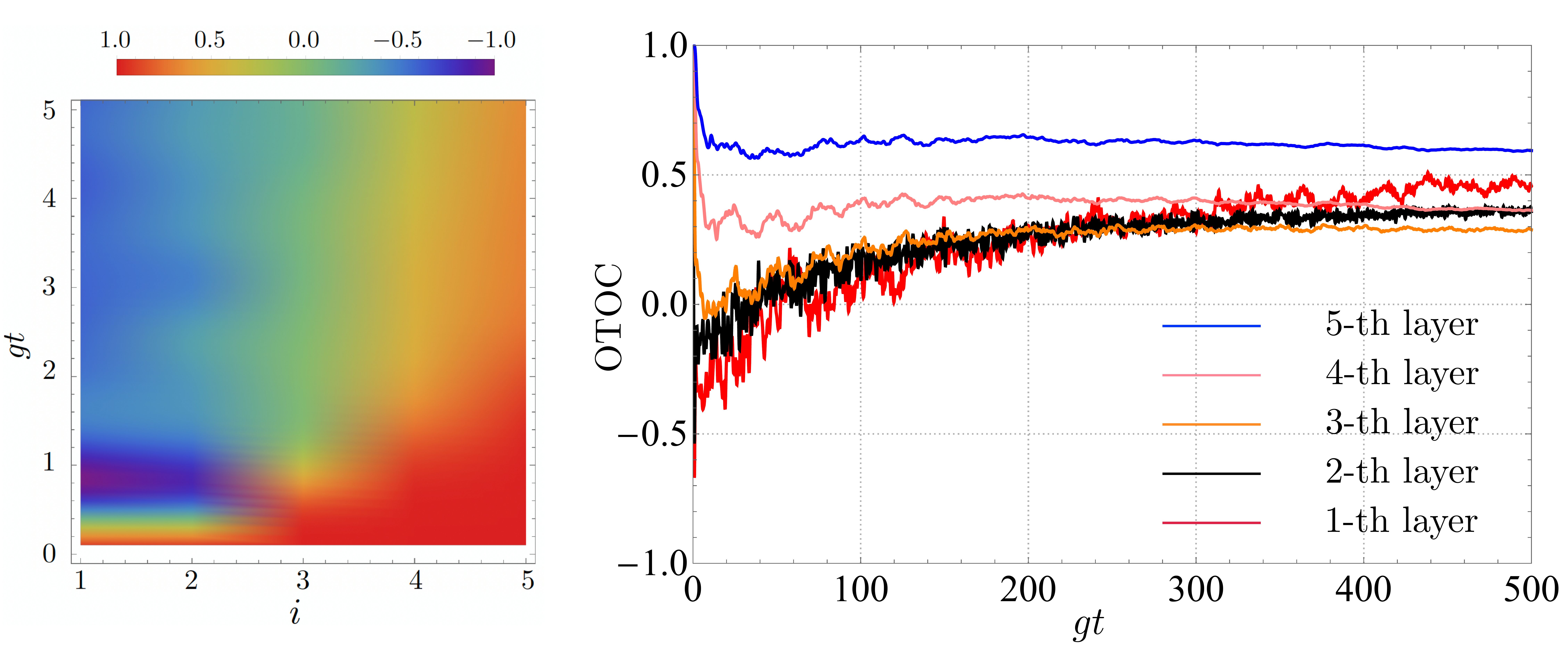}
\includegraphics[width=0.34\textwidth]{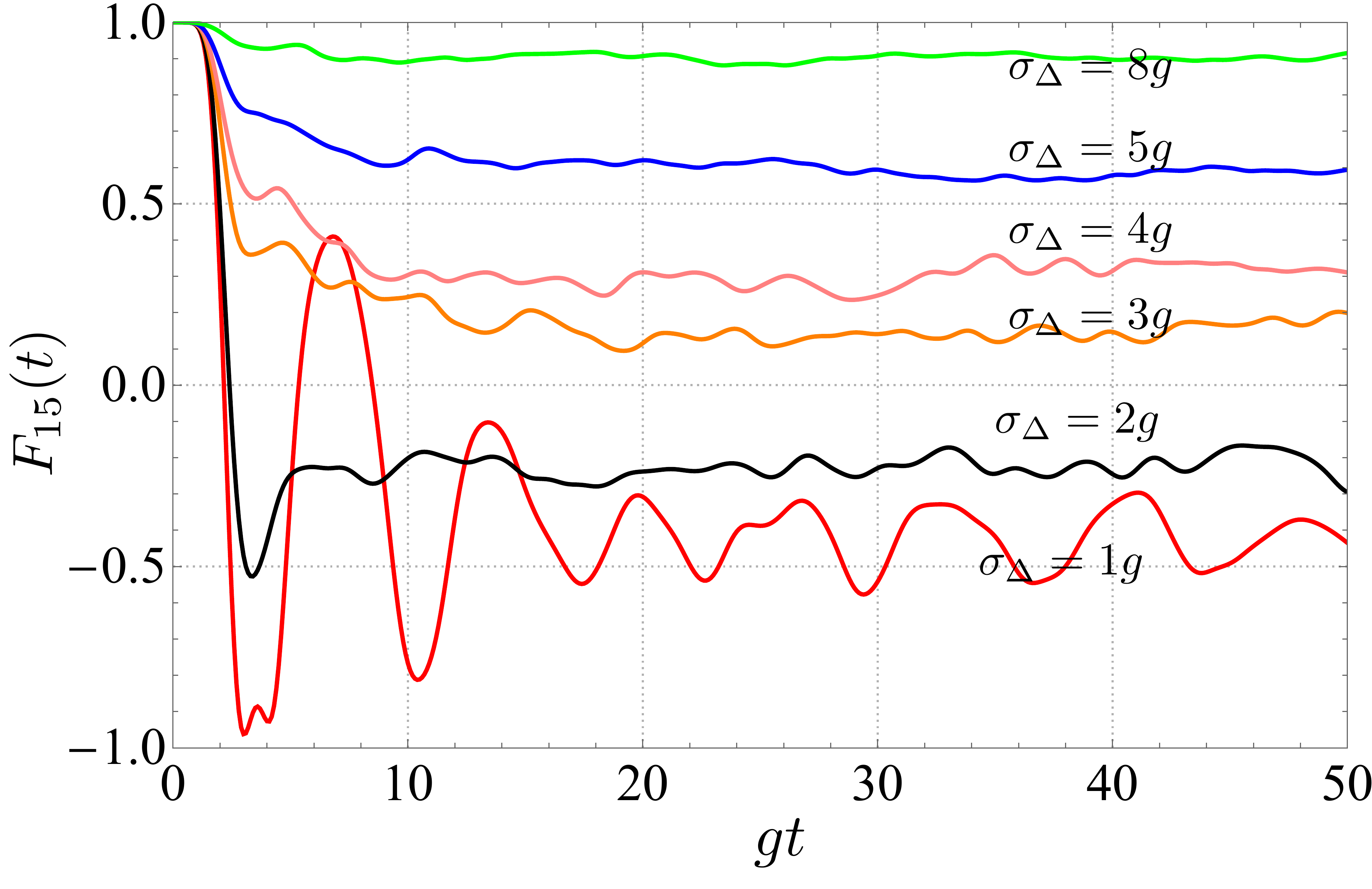}
\includegraphics[width=0.35\textwidth]{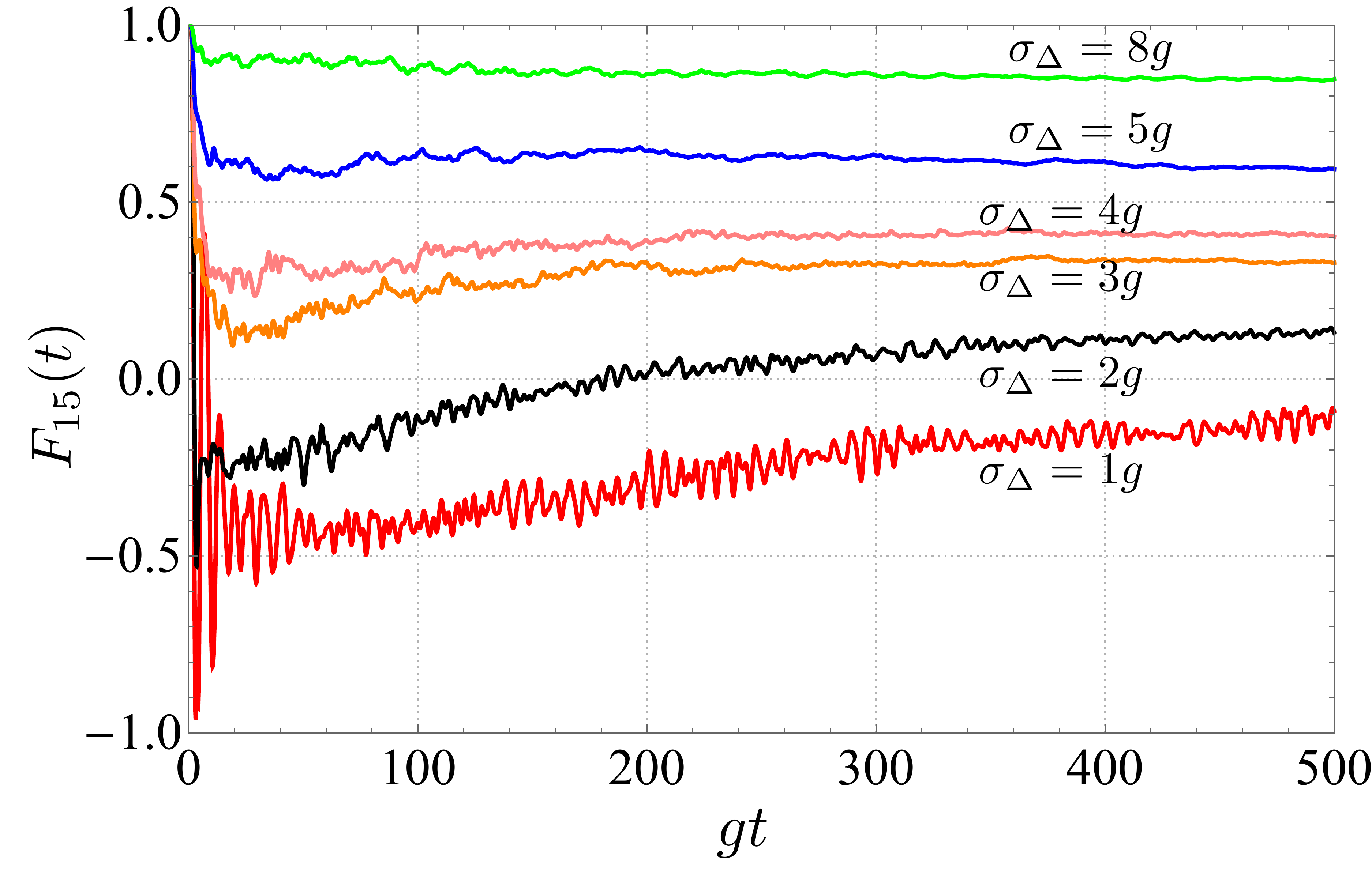}
\caption{OTOC evolution of the toy model with time-independent qubit level fluctuations among different sites, \emph{averaged over $10$ realisations of the randomness}. The upper left panel is the space-time evolution of OTOC on a short time scale with disorder strength $\sigma_\Delta=5g$, and it is clear that the quantum information propagation is blocked even at the early time.  The upper right panel is the OTOC evolution for different layers.  The mid/lower panel is the evolution of the averaged OTOC between the first qubit and a qubit in the fifth layer $F_{15}(t)$ with different disorder strengthes $\sigma_\Delta$ in the short/long time scale.  }\label{fig:localisation_average}
\end{figure}

Our toy model behaves differently, which can be understood as follows\,(see the lower panel of Fig.\,\ref{fig:irreversibility}). In the Anderson model, the hopping happens between neighbouring sites with a conserved occupation number. The hopping in our model accompanies the avalanche process, where the excitation numbers are not conserved. This creates an important difference which is crucial to understanding irreversibility:
For the forward process described by an interaction kernel as $\sim g\hat\sigma_{(-)}^{i}\hat\sigma_{(+)}^{i+1,j}\hat\sigma_{(+)}^{i+1,j+1}e^{-i\Delta\phi}$, i.e. annihilating the occupied state in the $i_{\rm th}$ layer while creating the double excitations of the $(i+1)_{\rm th}$ layer, the phase matching error is $\Delta\phi=(\omega_{i}-\omega_{i+1,j}-\omega_{i+1,j+1})t$. If we assume the probability for the phase-matching is $p$ at all interaction kernels, then the probability for this forward process to be phase-matched is $p_{\rm for}\sim p\times p_o$, where $p_o$ is the probability of an occupied site. The backward process described by $\sim g\hat\sigma_{(+)}^{i}\hat\sigma_{(-)}^{i+1,j}\hat\sigma_{(-)}^{i+1,j+1}e^{i\Delta\phi}$, is a process where the double excitation at the $(i+1)_{\rm th}$-layer annihilated while creating a single excitation at the $i_{\rm th}$-layer. In this case, the condition of doubly excited $(i+1)_{\rm th}$-layer can efficiently happen with the probability $p_{\rm back}\sim p\times p_o^2<p_{\rm for}$. Therefore the randomness in our model creates a weaker localisation of quantum information compared to the Anderson model. Furthermore, Fig.\,\ref{fig:localisation_average} demonstrates the OTOC evolution averaged over 20 realisations of random qubit levels of the network, where the wiggles in the Fig.\,\ref{fig:localisation} are suppressed. 

Fig.\,\ref{fig:localisation_average} shows how the evolution of the correlation between the first qubit and the qubit in the last layer\,(e.g. $F_{15}(t)$) depends on the disorder strength $\sigma_\Delta$. The results show that the $F_{15}(t)$ increases (or the correlation decreases) with the increasing of the disorder strength $\sigma_\Delta$, which is also a sign of the localisation.

\subsection{Temporally random qubit levels.} 
\begin{figure}[h]
\centering
\includegraphics[width=0.35\textwidth]{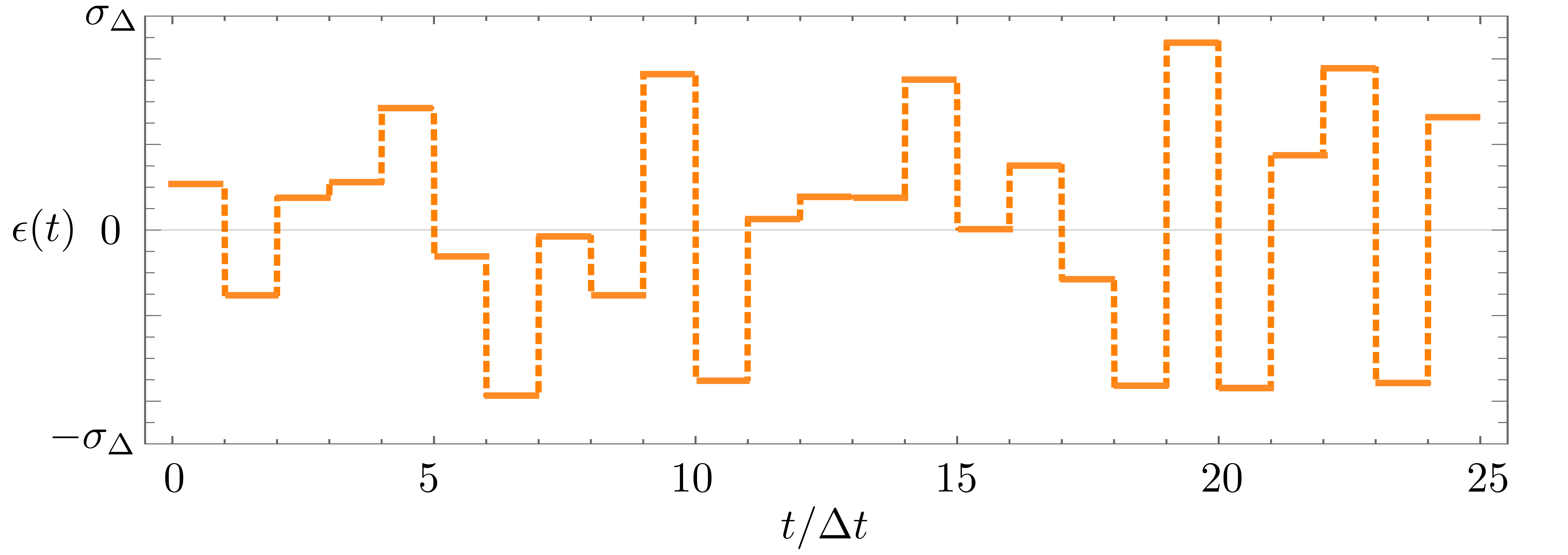}
\includegraphics[width=0.5\textwidth]{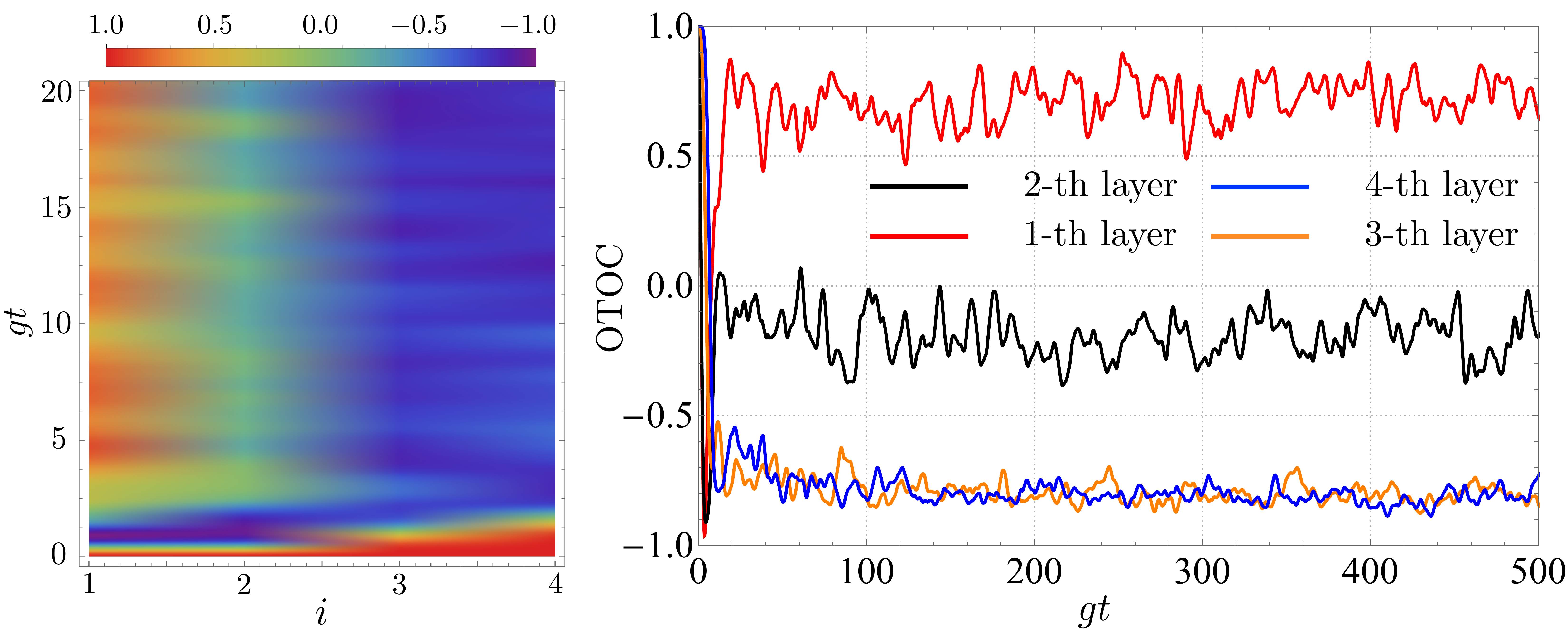}
\includegraphics[width=0.5\textwidth]{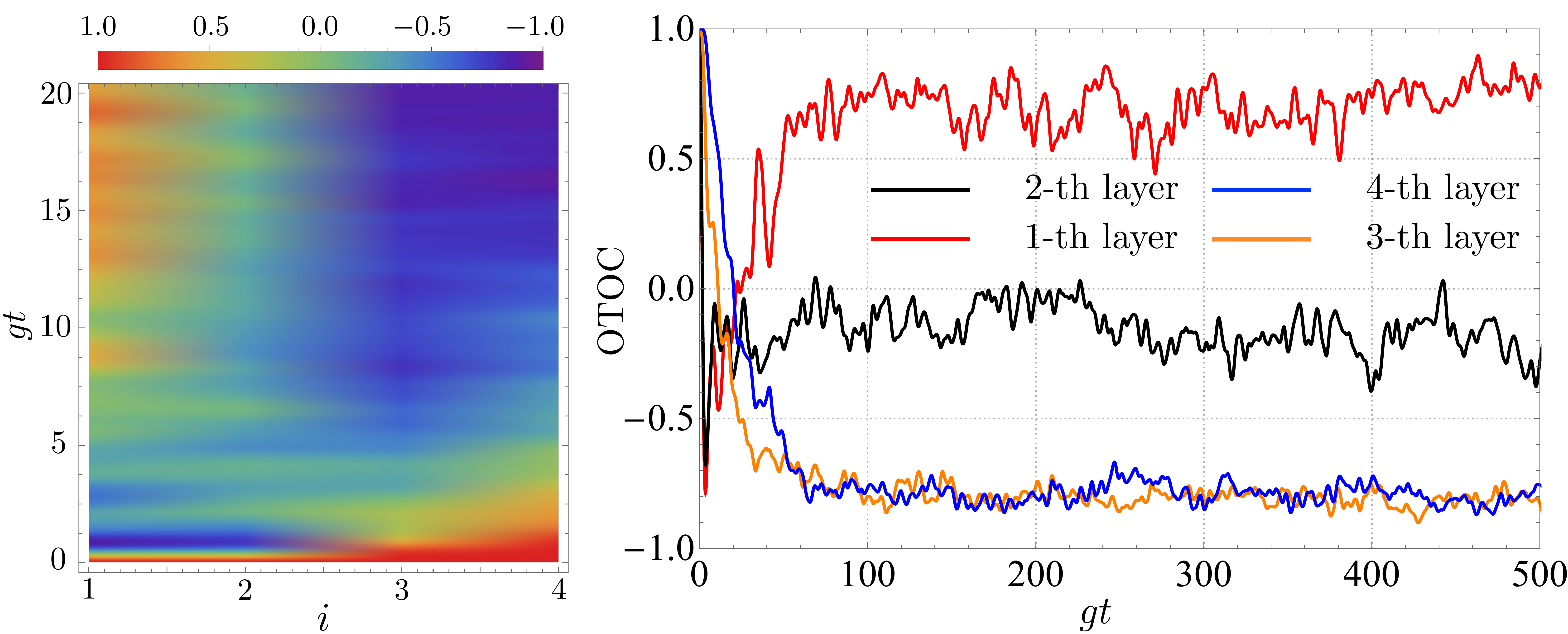}
\caption{Evolution of the averaged out-of-time correlator $\bar F_{1j}(t)$ of four layers quantum photo-detector toy model, where the $j$ labels the qubit located at the left boundary of each layer and the qubit level fluctuates both spatially and temporally. The result shown here is averaged over 10 different realisations. Upper panel: one example showing how we choose a realisation of the time-dependent disorder, in the units of time variations scale.  Middle panel: the color bar represents the value of the OTOC. The upper left/right figures shows the space-time evolution of the OTOC in the long/short time scale. Lower panel: The long-time evolution of OTOC when $j$ runs over different layers.}\label{fig:otoc_spacetime}
\end{figure}

The more realistic scenario is that the qubit energy levels are fluctuating in time and randomly distributed among different sites. The corresponding behaviour of the OTOC is plotted in Fig.\,\ref{fig:otoc_spacetime}. This time-dependent disorder breaks the localisation in the previous section and the quantum information and the excitations will accumulate on the last two layers at the later evolution stage.

In the interaction picture, the evolution operator of a typical interaction kernel (e.g. $A-B_1-B_2$) can be written as:
\begin{equation}\label{eq:random average}
\hat{U}_I(\tau, 0)={\rm exp}\left[-ig\int_0^\tau \left(e^{i\int_0^{\tilde \tau} \epsilon(\tau')d\tau'}\hat\sigma_{-}^{A}\hat\sigma_{+}^{B1}\hat\sigma_{+}^{B2}+h.c\right)d{\tilde \tau}\right],
\end{equation}
where $\epsilon (t)=\Delta_A(t)-\Delta_{B1}(t)-\Delta_{B2}(t)$, which is set to fluctuate around zero. For illustration, the distribution of the $\epsilon$ is set to be an uniform distribution within $[-\sigma_\Delta,+\sigma_\Delta]$ and the time-dependence manifests as a revision of all the $\epsilon_i$ in the network by every time interval $\Delta t$. In the numerical simulation, we reduce the integration to a discrete summation as shown in the upper panel of Fig.\,\ref{fig:otoc_spacetime}, and 
the simulation result is plotted in Fig.\,\ref{fig:otoc_spacetime} with the same disorder strength $\sigma_{\Delta}=5g$ but different time variation interval $\Delta t=2/g$ and $\Delta t=0.2/g$. 
For the small $\Delta t$ and disorder strength satisfying $\Delta t\sigma_\Delta\ll 1$, the influence of the time-dependent phase mismatch in Eq.\,\ref{eq:random average} is very weak, therefore the system will still exhibits recurrence behaviour as in the ideal case. On the contrary, for the larger $\Delta t$ and disorder strength satisfying $\Delta t\sigma_\Delta\geq 1$, the recurrence phenomenon disappears and the information will be trapped on the last two layers, since the quantum information flow backward encounters a different realisation of the disorder.   In the limit $\Delta t\to \infty$, the evolution will return to the previous model with time-independent disorders. Note that the time variation scale difference mainly affect the evolution at the relatively early stage, at the long time tail the evolution of OTOCs converge to a stable value, since the quantum information has been strongly scrambled in the system.


\begin{figure}[h]
\centering
\includegraphics[width=0.39\textwidth]{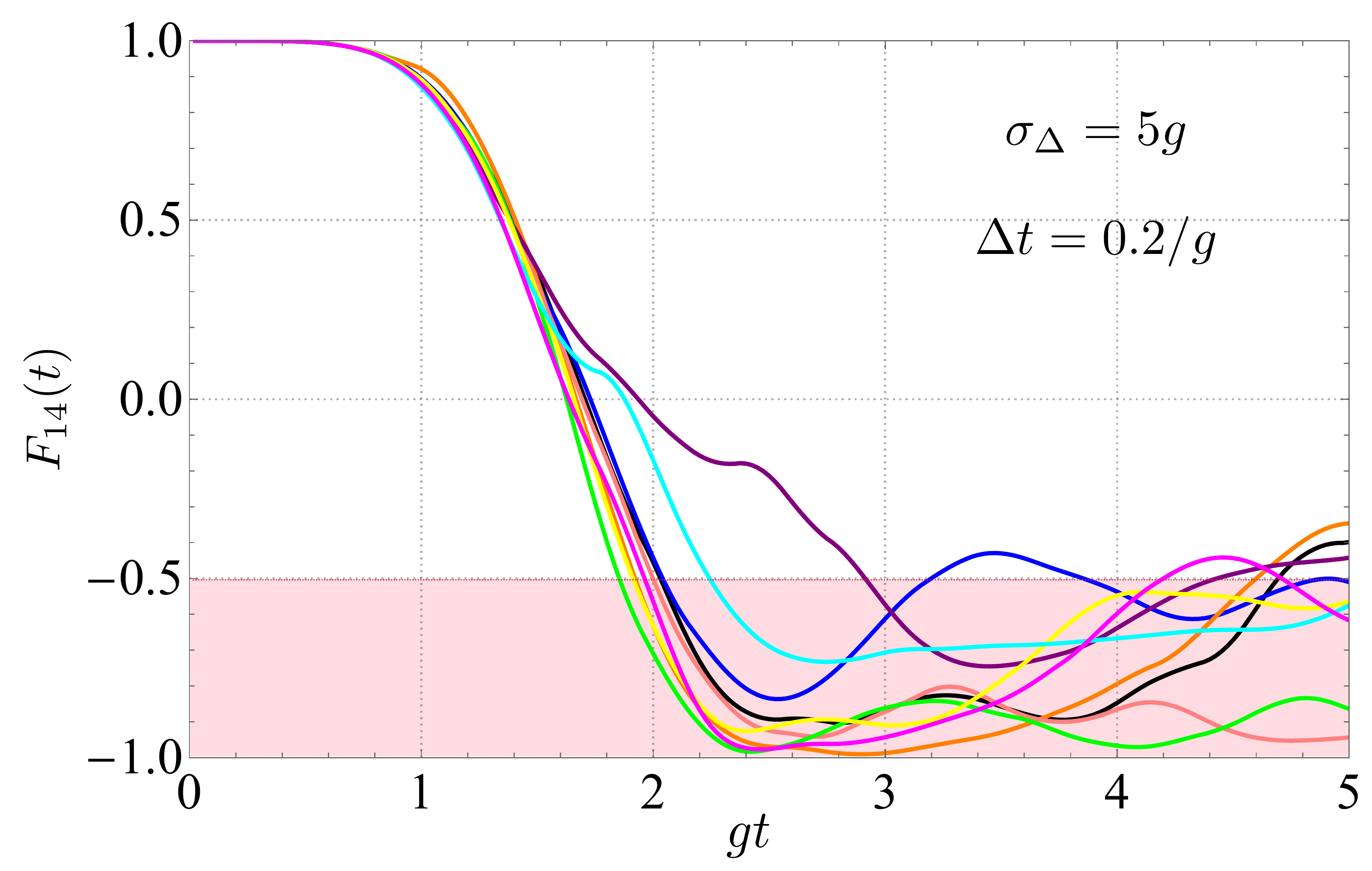}
\includegraphics[width=0.4\textwidth]{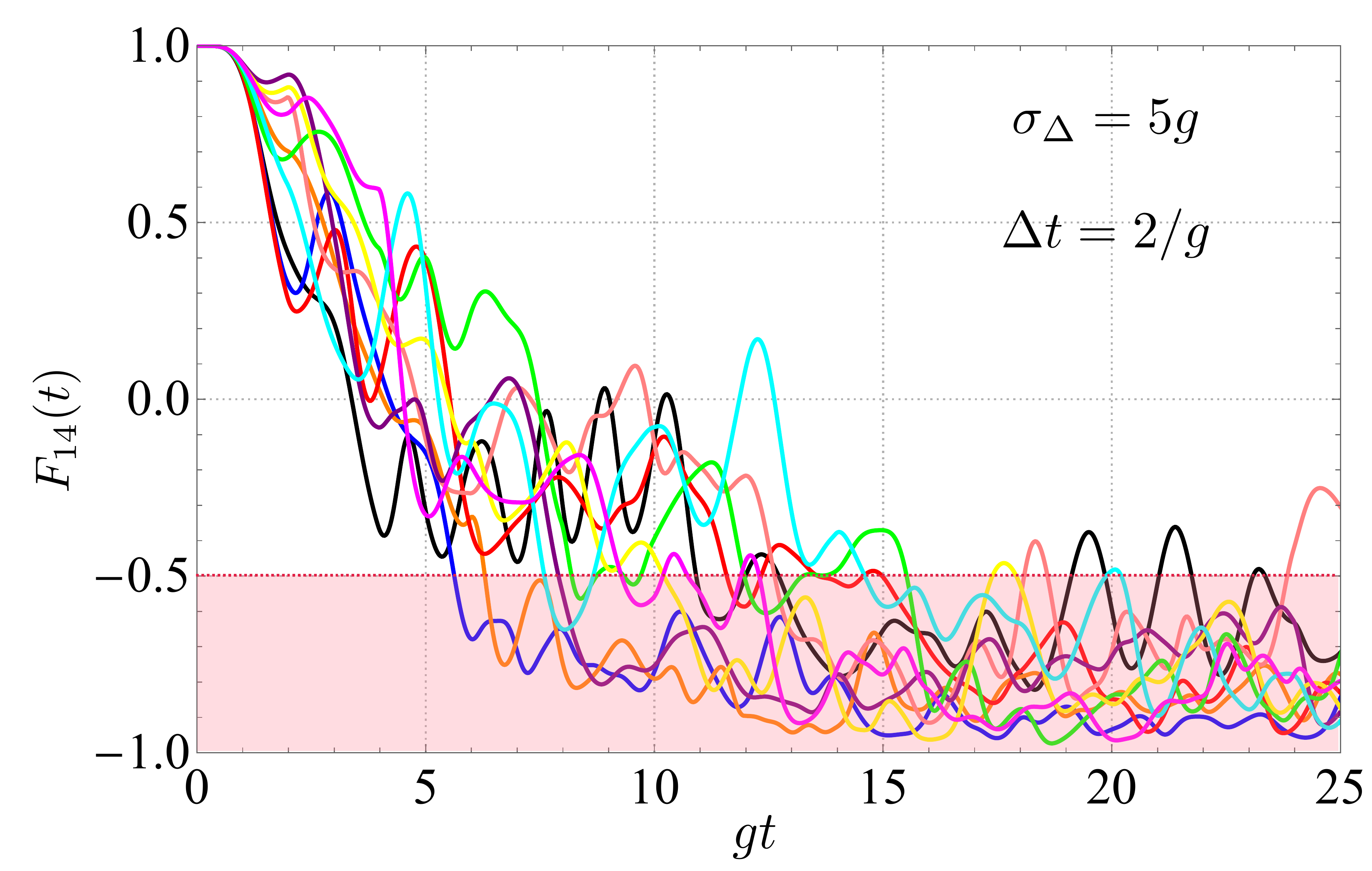}
\caption{Evolution of the out-of-time correlator $\bar F_{14}(t)$ of four layers quantum photo-detector toy model, for 10 different realisations of time-dependent random detuning $\epsilon(t)$. The upper and lower panels correspond to different time variation scale $\Delta t$. We set the $F_{15}<-0.5$ as the threshold for the arrival of quantum information on the last layer of the network. Note that the arrival time  of the quantum information  to the last layer for different realisations also randomly distributes. We only show the early evolution stage, which is less affected by the backward propagating information.}\label{fig:arrival_time}
\end{figure}

It is also interesting to investigate the (un-averaged) OTOC for different realisations, which is shown in Fig.\,\ref{fig:arrival_time}. In particular, we want to explore how the time-dependent disorder in the network affect the arrival of the initial quantum information on the first qubit site to the last layer, with a significantly amplified occupation number by the avalanche process. As we mentioned in the beginning of this section, a more negative OTOC means a stronger correlation of the operators, this leads us to set the $F_{1j}\leq -0.5$ as a threshold to judge the arrival of quantum information on the $j_{\rm th}$ layer. To avoid the recurrence effect by the information propagate backward from the last layer,  Fig.\,\ref{fig:arrival_time} only shows the early stage of the OTOC evolution and the results indicates that there will be a stronger scattering of the arrival time when the time variation scale is larger. 

\section{Evolution of the Holevo information}\label{sec:5}
As pointed out in\,\cite{Xu2022}, treating the squared commutator $C_{ij}(t)$ or the OTOC $F_{ij}(t)$ as an indicator of the quantum information propagation when the initial state is a simple maximumly mixed state. The relation between OTOC and the mutual information is not well-established when the initial state is not maximumly mixed. Therefore, in this section, we further investigate the quantum information evolution using the Holevo information, which is used to characterise the accessible information (or maximum mutual information) between two separate agents in a complex system.

\begin{figure}[h]
\centering
\includegraphics[width=0.5\textwidth]{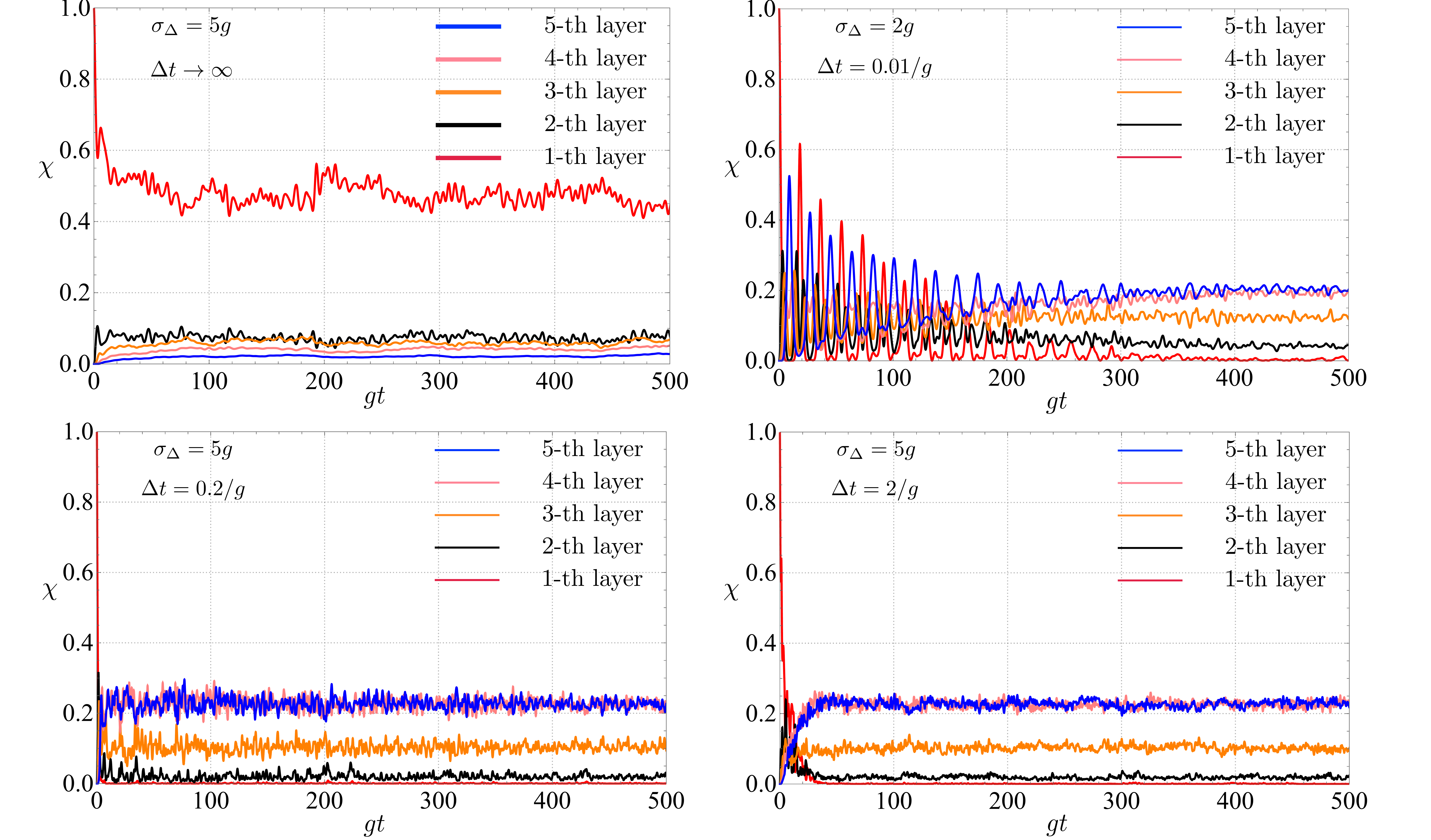}
\caption{Evolution of the Holevo information $\chi_{ij}(t)$ of a five layer quantum photon-detector toy model, where the $j$ takes the qubit located at the left boundary of each layer. The upper panel describes the Holevo information when the qubit level fluctuation is time independent with $\sigma_\Delta=5g$, while the lower panel describes the case when the qubit level fluctuation is time-dependent. }\label{fig:holevo}
\end{figure}

The formula of Holevo information is\,\cite{Dong2022}:
\begin{equation}
\chi=S(\sum_{i=1}^n p_i\hat\rho_i)-\sum_{i=1}^n p_iS(\hat\rho_i),
\end{equation}
where $S(\hat\rho)$ is the von Neumann entanglement entropy with definition $S(\hat\rho)=-{\rm Tr}(\hat\rho{\rm Log}\hat\rho)$. Holevo information can be interpreted as the measurement of the difference of different quantum states in $\{\hat\rho_1,\hat\rho_2,...,\hat\rho_n\}$. As a simple example, if $p_1=p_2=1/2$, then $\chi=1$ for $\hat\rho_1=|0\rangle\langle 0|$, $\hat\rho_2=|1\rangle\langle 1|$ and $\chi=0$ for $\hat\rho_1=\hat\rho_2$. Now we consider a local operator $\hat{O}_i$ acting on $i_{\rm th}$ qubit and the density matrices $\hat\rho$ and $\hat\rho^{\prime}$ are defined to describe the state evolution from the initial network state $|\psi\rangle$ and $\hat{O}_i|\psi\rangle$ respectively. To characterise how this local operation on the $i_{\rm th}$-site affects the $j_{\rm th}$-site, we define the $\tilde\rho_j$ and $\tilde \rho'_j$ as the reduced density matrix of the $j_{\rm th}$-qubit before and after the local operation, respectively. In this case, the Holevo information can be written as:
\begin{equation}
\chi_{ij}(t)=S\left[\frac{\tilde\rho_j(t)+\tilde\rho_j^{\prime}(t)}{2}\right]-\frac{S(\tilde\rho_j(t))+S(\tilde\rho^{\prime}_j(t))}{2},
\end{equation}
of which the evolution is plotted in Fig.\,\ref{fig:holevo}, and exhibits the same behaviour as the OTOC.

\section{Connections to quantum-state Collapse, Conceptual Issues and Outlook}

The key issue with the quantum measurement process is that identical initial states lead to different phenomenological outcomes.  Unless we take the relative-state (or many-world) point of view, we will need to provide {\it additional information} in order to explain why this particular outcomes is achieved at this particular trial of the ``same experiment''. 

\subsection{Connection and Conceptual Issues}

From numerical simulations, we have seen that: (i) the values of $g_j$ controls the time at which the avalanche toward lower layers takes place, (ii) if the number of layers can be extended, information will tend not to flow back to higher layers. This supports the notion that internal mechanisms of the detector supplies that additional information.  The complexity of the device also prevents further observers from modifying the result of quantum measurement.  However, as one attempts further to solve the ``measurement problem'' this way, one encounters two serious challenges: the non-locality problem, and the Schroedinger-Cat problem.  

In the non-locality problem, if we use two of our detectors to probe entangled photons at space-like separations, then each of their internal states, i.e., the $g_j$'s will have to influence the outcome of the other experiment.  In this way, we seem to have constructed a superluminal communcations device.  In the cat problem, if we prepare the atom at a superposition state $\alpha|g\rangle + \beta |e\rangle$, then from quantum mechanics we already have a superposition of two states, one with avalanche, the other without.  For this reason, it is possible that while this many-body model resembles the quantum collapse, it does not in fact explain it.  Nevertheless, we shall try to outline some loopholes below. 

\subsection{Loopholes}

  For the non-locality problem,  superluminal communication requires the observer to be in total control of the device, which is not practical.  If the observer is able to divulge the measurement result, then additional influence from the same channel can also affect the measurement result.  Furthermore, if the observer is then going to use the measurement result to create logical inconsistencies, back-action from the follow-up device will not allow inconsistencies to take place.  This has been argued for in the context of closed time-like curves by Novikov~\cite{novikov1992time}, Friedman et al.~\cite{friedman1990cauchy} in the 1990s, and more recently by Tobar and Costa~\cite{tobar2020reversible}. 

For the cat problem, we need to realize that the state-preparation process is just another measurement process that took place before the ``official'' measurement starts.  In this way, a ``superposition state'' for the system being prepared is really a collection of possible outcomes in the preparation devices, instead of a superposition state for the joint system that consist of the preparation and measurement devices.  A more fundamental treatment of the preparation-measurement procedure should incorporate both the preparation and measurement devices, and in this situation the cat paradox may not arise. 

\subsection{Outlook}

Finally, to move toward understanding realistic quantum measurement processes, we need to model the entire environment of the quantum system.  Consider a single atom that is prepared into an excited state and radiates a photon, which can either be detected by one of several measurement detectors, or leave toward infinity.   The atom also had to be monitored by preparation detectors, whose signal must indicate that the atom is present and at the correct initial quantum state.    One will need to elaborate how the $g_j$'s within the measuring devices, as well as those within the preparation devices, together reproduces the spontaneous emission rate of the atom, as well as the correct branching ratio toward each of the measuremnt detectors. This seems highly non-trivial; from our numerical simulations, it seems that if $g_j$'s are appropriately tuned, then the ``collapse process'' can be accelerated indefinitely.   Here we should be reminded of Wheeler and Feynman's work on absorber theory of radiation\,\cite{Wheeler1945}, in which the spontaneous emission rate is always recovered even in presence of absorbers that seem to have high susceptibility to radiation. One way to understand this is that radiation damping within the device will not allow it to respond to the emitter's radiation faster than rate that will make it emit more than the spontaneous emission rate.

\section{Conclusion and Discussions}\label{sec:6}
In this paper, we investigated the quantum information scrambling in a qubit-network toy model of the quantum measurement process, which resembles the typical photo-detector that generates the macroscopic measurable photo-current through the avalanche process. Our system is studied based on a full-unitary evolution model, where the environmental effect is taken into account through the fluctuation of the qubit levels. The eigenvalue level spacing statistics of our model in the reduced Hilbert space is numerically computed and the result exhibits a Poission distribution, which indicates that our system is a integrable system in the reduced Hilbert space. The evolution of the out-of-time correlator and the Holevo information is simulated in different physical scenarios. We found that there is localisation of the quantum information in our system with time-independent disorder, which is similar to the Anderson localisation in condensed matter physics. When the disorder becomes time-dependent, our simulation shows that the localisation will disappear and the quantum information finally would be kept in the last several layers in our model.

The interpretation of the quantum measurement process is a long-standing issue at the foundation of physics. One key question is: where does the unitary process end, and the device start, which is related to the interaction between, for example, the photon and the measurement device.
Even though the device is highly complex, it still falls into the current scope of quantum many-body physics.  Furthermore,  the energy scale of most quantum measurement processes is well within the typical energy scale of quantum electrodynamics, which makes it possible that no new physics, hidden at different energy scales (e.g. predicted by Bohm as ``sub-quantum" structure in\,\cite{Bohm1963}), play a critical role to this process --- although it  had also been argued that gravity may play a role in this process\,\cite{Penrose1996}.

The modern development of quantum technologies, in particular quantum information and computing technology, provide the possibility to simulate the details of the quantum measurement process, by customer-designing a table-top quantum measurement apparatus, observing and even manipulating the details of the measurement process. Though far from providing an answer to the quantum measurement problem, the purpose of this work is a hope to trigger the theoretical and experimental research in this direction in the current and future quantum era, on the various quantum simulator platforms such as neutral atoms, superconducting qubits and ion-trap. This could be a step forward to understanding the mystery of quantum measurement or ``wave function collapse".

\acknowledgements
Y.L.\ and Y.M.\ thank Yinghai Wu, Xiaopeng Li and Daiqin Su for very helpful discussions. Y.M.\ also thank Shun Wang for the discussion and his encouragement. Y.M.\ thanks Jingtao L{\"u} for the conversation about thermalisation in solid-state physics.  Y.M.\ is supported by the start-up fund provided by Huazhong University of Science and Technology. H.M.\ is supported by the State Key Laboratory of Low Dimensional Quantum Physics and the start-up fund from Tsinghua University. Y.C.\ is supported by Simons Foundation (Award Number 568762) .
\appendix


\bibliographystyle{unsrt}
\bibliography{reference}

\end{document}